\documentclass[12pt]{article}
\usepackage{epsf,epsfig,float,amssymb,latexsym,amsmath,amsthm,fancyhdr,pst-all}
\usepackage{graphics,psfrag,citesort}
\newcommand{\xpc}{\%}
\newcommand{\om}{\Omega}
\newcommand{\la}{\langle}
\newcommand{\ra}{\rangle}

\newcommand{\be}{\begin{equation}}
\newcommand{\ee}{\end{equation}}
\newcommand{\ba}{\begin{eqnarray}}
\newcommand{\ea}{\end{eqnarray}}

\newcommand{\nn}{\nonumber}

\newcommand{\vs}{\vspace{-0.20cm}}

\begin{document}

\thispagestyle{empty}

\vspace{2cm}

\begin{center}
{\Large{\bf Two photons  into $\pi^0\pi^0$}}
\end{center}
\vspace{.5cm}

\begin{center}
{\large J. A. Oller, L. Roca }
\end{center}

\begin{center}
{\it {\it Departamento de F\'{\i}sica. Universidad de Murcia. E-30071,
Murcia. Spain.\\
{\small oller@um.es~,~luisroca@um.es~}}}
\end{center}
\vspace{1cm}

\begin{abstract}
\noindent
We perform a theoretical 
study based on dispersion relations of the reaction $\gamma\gamma\to \pi^0\pi^0$ emphasizing
 the low energy region. 
We discuss how the $f_0(980)$ signal emerges in $\gamma\gamma\to \pi\pi$ within the dispersive approach 
and how this fixes to a large extent the phase of the isoscalar
 S-wave $\gamma\gamma\to \pi\pi$ amplitude above the $K\bar{K} $ threshold. 
  This allows us to make sharper predictions for the 
cross section at lower  energies and  our results could then be used to distinguish between different 
$\pi\pi$ isoscalar S-wave parameterizations with the advent of new precise data on $\gamma\gamma\to\pi^0\pi^0$. 
  We compare our dispersive approach with an updated calculation 
  employing Unitary Chiral Perturbation Theory (U$\chi$PT). 
 We also pay special attention to the role played by 
   the $\sigma$ resonance in $\gamma\gamma\to\pi\pi$
 and calculate its coupling and width to $\gamma\gamma$,
 for which we obtain $\Gamma(\sigma\to\gamma\gamma)=(1.68\pm 0.15)~$KeV.
\end{abstract}

\vspace{2cm}


\newpage

\section{Introduction}
\label{sec:intro}
\def\theequation{\arabic{section}.\arabic{equation}}
\setcounter{equation}{0}

The study of the reaction $\gamma \gamma \to \pi^0\pi^0$ offers the
possibility of having  two neutral pions in the final state as a
two body hadronic system. In this way, the interactions among pions can be
unraveled more cleanly than in other processes where three or more
hadrons appear in the final state. In this respect, the  $K_{e4}^{00}$ 
decays also share this property but limited to energies below the kaon
mass. Furthermore,  the two photon fusion also offers the
interesting property of coupling mostly to the charged components
of the hadronic states, while strong interactions are blind to
that. Therefore, the study of $\gamma\gamma$ fusion into two
 hadrons displays new information on the inner constituents, beyond
that obtained in direct strong scattering data. 
 In addition,  $\gamma\gamma\to \pi^0\pi^0$ has no either Born term, 
shown in fig.\ref{born} for charged mesons, so that final state interactions 
dominate this reaction.  In this respect,
one can think  of the interest of precise data on this reaction to
 distinguish between present low energy parameterizations of the $\pi\pi$ isospin ($I$) 0 S-wave, 
 to further study the nature of the $\sigma$ resonance 
 as well as that of the $f_0(980)$, and having another way
to constraint their pole positions. The presence of the $\sigma$ and its tight relation to 
 chiral symmetry was clearly established from those 
non-perturbative methods  \cite{npa,nd,iamcc,mrios,hannah} that resum unitarity in the series of Chiral 
 Perturbation Theory ($\chi$PT) \cite{wein,glchpt}. These methods  were the first to show that 
the presence of a $\sigma$ pole is not incompatible with the non-linear $\sigma$ models. They 
also showed  a clear convergence in 
the $\sigma$ pole position when using either lowest order  or one or two loop $\chi$PT 
in the resummation, this is 
discussed in detail in ref.\cite{hannah}.
 Recently, its associated pole was determined by solving Roy equations together with constraints 
 from $\chi$PT to two loops \cite{caprini}. Other recent determinations are refs.\cite{pelaysig,alba}.

The reaction $\gamma \gamma\to \pi^0\pi^0$ is also an 
interesting ground test for $\chi$PT \cite{wein,glchpt}, 
since at lowest order is zero and
at next-to-leading order  (one loop) it is a prediction free of any
counterterm \cite{bc88,dhl87}. The one loop  $\chi PT$ calculations
were performed before the advent of data but the prediction just above threshold
departs  rapidly from experiment \cite{crystal}   
 and only the order of magnitude was rightly
foreseen. A two loop $\chi$PT calculation was done in
refs.\cite{bgs94,gis05}, where the agreement with data \cite{crystal}
at low energies was improved. The three counterterms that appeared at
${\cal O}(p^6)$  were fixed by the resonance saturation hypothesis \cite{colla}.
Other approaches supplying higher orders to one loop $\chi PT$  by
taking into account unitarity \cite{dp93} plus also crossed
exchange resonances \cite{dh93,oo98} followed. Ref.\cite{oo98} is a  dynamical approach, 
with higher predictive power than dispersion relations (which make use of extensive experimental input 
from other related processes), though more model dependent. Ref.\cite{oo98} 
 is a Unitary $\chi$PT (U$\chi$PT) calculation in production processes
\cite{oo98,phi1,hyper,sfm,vfm,anke} and 
was able to simultaneously describe $\gamma\gamma \to \pi^0\pi^0$,
$\pi^+\pi^-$, $\eta \pi^0$, $K^+K^-$ and $K^0\bar{K}^0$ from
threshold up to rather high energies, $s^{1/2}\lesssim 1.5$~GeV. This approach 
was also used in ref.\cite{rocapela} to study the $\eta\to\pi^0\gamma\gamma$ 
decay.

Recently, ref.\cite{penprl}, making use of the approach of
refs.\cite{penmorgan,penanegra}, focused on the low
energy part of the reaction $\gamma\gamma\to\pi^0\pi^0$. 
Ref.\cite{penprl} emphasizes the calculation of the $\gamma\gamma$ 
 coupling of the $\sigma$ as a way to deepen in the 
understanding of its nature.
 This approach was extended in ref.\cite{orsletter} so
 as to largely remove the sensitivity for lower energies, $\sqrt{s}\lesssim 0.8~$GeV, on the 
 uncertainty in the not precisely known phase of the $I=0$ S-wave 
 $\gamma\gamma\to\pi\pi$ amplitude 
 above the $K\bar{K}$ threshold. The novelty was to include a further subtraction in the dispersion
 relation for the $I=0$ S-wave $\gamma\gamma\to\pi\pi$, together with an extra constraint 
 to fix the additional subtraction constant. This was motivated by the use of an
 improved $I=0$ S-wave Omn\`es function in the lines of ref.\cite{or07}. 
 We shall here extend the findings of ref.\cite{orsletter} and discuss in detail 
 how the $f_0(980)$ peak, clearly seen recently in $\gamma\gamma\to \pi^+\pi^-$ \cite{mori},  
can be generated within  the dispersive method of ref.\cite{orsletter}. We also show 
how this can be used to fix rather accurately the phase of the $I=0$ 
S-wave $\gamma\gamma\to\pi\pi$  amplitude above the 
 $K\bar{K}$ threshold. As a result, the remaining ambiguity in this phase by $\pi$ 
 in ref.\cite{orsletter}  is removed. 
  Interestingly, this allows to sharpen the prediction of 
 the $\gamma\gamma\to\pi^0\pi^0$
 cross section for $\sqrt{s}\lesssim 0.8$~GeV, up to the onset of D-waves, as compared with 
 ref.\cite{orsletter}. This could then be used 
to constraint further different parameterizations of the low energy $\pi\pi$ $I=0$ S-wave.   
 We also offer explicit expressions for several axial-vector and 
vector crossed exchange contributions to $\gamma\gamma\to \pi\pi$, that are used in the present 
analysis and that of ref.\cite{orsletter}. The  Unitary $\chi$PT (U$\chi$PT) calculation of ref.\cite{oo98} on 
$\gamma\gamma\to \pi\pi$ will be reviewed and extended, so as to incorporate the same 
 vector and axial-vector 
exchange contributions as in the dispersive analysis. 
 We finally discuss the $\sigma$ coupling to $\gamma\gamma$ and determine  
 $\Gamma(\sigma\to \gamma\gamma)=(1.68\pm 0.15)~$KeV.  Other 
papers devoted to the calculation of the 
two photon decay widths of hadronic resonances are 
 \cite{rosner,isgur,barnes,chano,narison,closebarn,munz,achasov,krewald,mexico,hanhart,narison2}.

The content of the paper is as follows. In section \ref{sec:dis} we
 discuss the dispersive method of
 ref.\cite{penmorgan} and its extension by ref.\cite{orsletter}
 for calculating  $\gamma\gamma\to \pi\pi$. Here we pay attention to the intrinsic difficulties  
 of the approach of 
 refs.\cite{penmorgan,penanegra,penprl} to generate the $f_0(980)$ peak in $ \gamma\gamma\to \pi\pi$ and 
 how this is overcome by the approach of ref.\cite{orsletter}. In this way, 
  one can use this extra information to improve the calculation at lower energies. In section \ref{sec:dyn} we discuss
 and revise the U$\chi$PT approach of ref.\cite{oo98}. We also give in this section 
several formulae  corresponding to the Born terms and 
exchanges of vector and axial-vector resonances to be used as input in both approaches. 
 The resulting cross section $\sigma(\gamma\gamma\to
\pi^0\pi^0)$ is discussed in section \ref{sec:resul}. The  $\gamma\gamma$
 decay width of the $\sigma$ resonance and its coupling to $\gamma\gamma$ is considered 
 in section \ref{sec:sigwidth}. Conclusions are elaborated in section \ref{sec:conclu}.

\section{Dispersive approach to $\gamma\gamma\to \pi\pi$}
\label{sec:dis}
\def\theequation{\arabic{section}.\arabic{equation}}
\setcounter{equation}{0}

 We first review the approach of refs.\cite{penprl,penmorgan,penanegra} to
establish a dispersion relation to calculate the  two photon
amplitude to $\pi\pi$. This approach has 
 a large uncertainty above $\sqrt{s}\gtrsim 0.5$~GeV, which dramatically rises with energy  such that for 
 $\sqrt{s}=0.5,~ 0.55,~ 0.6$ and $0.65$~GeV
 it is $20,~45,~92$ and $200\,\xpc$, see fig.3 of ref.\cite{penprl}. 
 This uncertainty was largely reduced by the extended method of ref.\cite{orsletter}. We discuss here
 how this method allows to disentangle also the role of the $f_0(980)$ resonance in $\gamma\gamma\to \pi\pi$ 
 and the way that this can be 
used to sharpen further the results of ref.\cite{orsletter}.
 
%
%
\subsection{Dispersion relation and Omn\`es function} 
\label{subsec:dis}

\begin{figure}[ht]
\psfrag{R}{$P^+$}
\psfrag{Q}{$P^-$}
\centerline{\epsfig{file=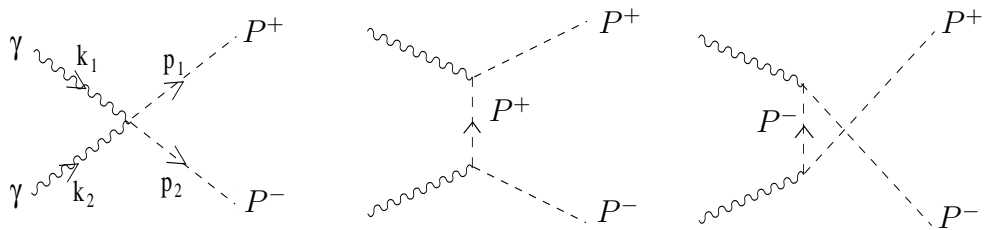,height=1.1in,width=5.0in,angle=0}}
\vspace{0.2cm}
\caption[pilf]{\protect \small
 Born term contribution to $\gamma(k_1) \gamma(k_2)\to P^+(p_1) P^-(p_2)$.  
\label{born}}
\end{figure} 

Due to the null charge of the $\pi^0$ there is no Born term, shown in fig.\ref{born},  for $\gamma\gamma \to 
\pi^0\pi^0$. As remarked in ref.\cite{penanegra}, 
one then expects that only the S-wave would be the important partial wave 
at low energies, $\sqrt{s}\lesssim 0.6-0.7$~GeV. For $\gamma\gamma\to \pi^+\pi^-$, where there is a Born
term because of the exchange of charged pions, the D-waves have a relevant contribution 
already at surprisingly low energies due to the smallness of the pion mass. 
 In the following, we shall restrict ourselves to the S-wave contribution to 
$\gamma\gamma\to\pi\pi$, being particularly interested in the low energy region. Explicit 
 calculations \cite{oo98} show that
 the D-wave contribution in the $\gamma\gamma\to\pi^0\pi^0$ cross section  
 at $\sqrt{s}\simeq 0.65$~GeV amounts less than a $10\xpc$, and this 
 rapidly decreases for lower energies.
Let us consider the S-wave amplitude $\gamma\gamma\to(\pi\pi)_I$,
$F_I(s)$, where the two pions have definite $I=0$ or 2. Notice that charge conjugation of the 
 photons is even and this, together with Bose symmetry, excludes odd orbital angular momentum 
 (and hence odd isospins) for the two pions. 
 The function $F_I(s)$ on the complex $s-$plane is analytic except for two cuts, 
 the unitarity one for $s\geq 4m_\pi^2$ and the left hand cut for $s\leq 0$, with 
 $m_\pi$ the pion mass. Let us 
denote by $L_I(s)$ the complete left hand cut contribution of $F_I(s)$. Then the
 function $F_I(s)-L_I(s)$, by construction, has only right hand cut. 

\begin{figure}[ht]
\centerline{\epsfig{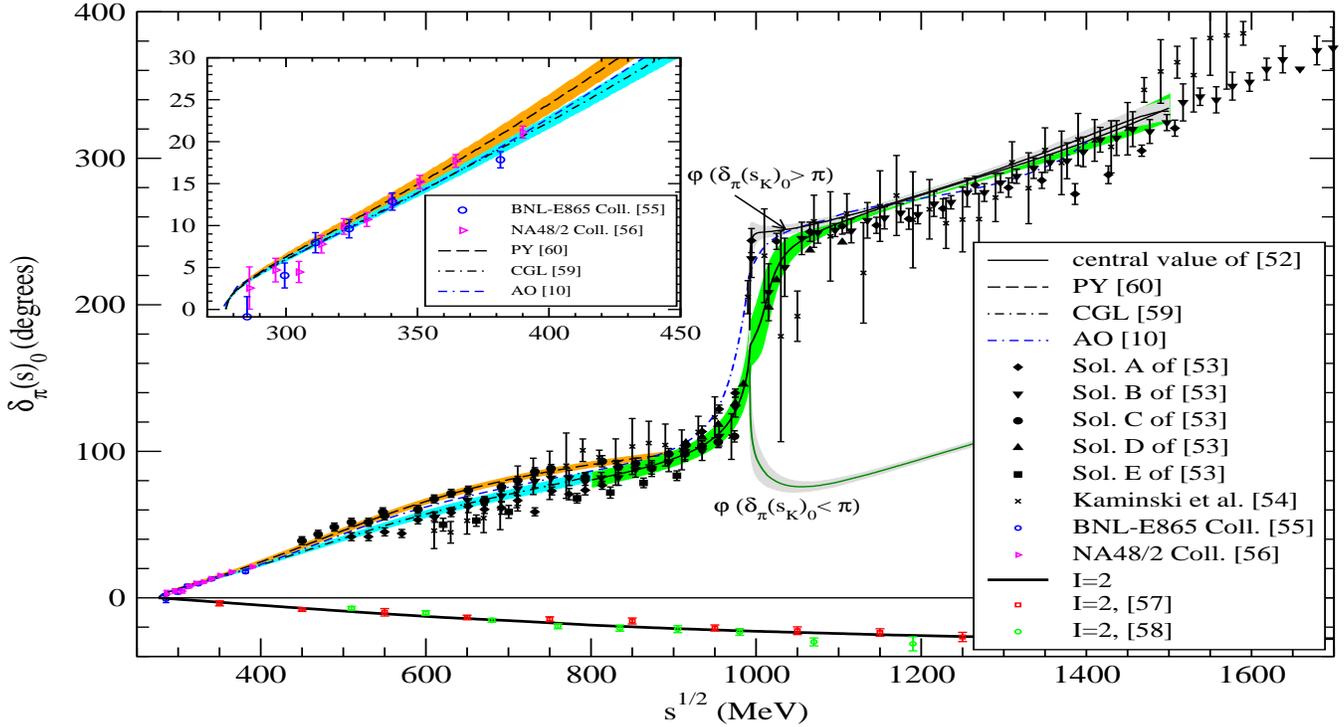}}
\vspace{0.2cm}
\caption[pilf]{\protect \small
The phase shifts $\delta_\pi(s)_0$ and 
$\delta_\pi(s)_2$ and  the phase $\varphi(s)$.
 Experimental data are from refs.\cite{hyams,kaminski,grayer,bnl,na48} for $I=0$ and 
 refs.\cite{losty,hoog} for $I=2$. The insert is the comparison of 
  CGL \cite{cgl}, dot-dashed line, PY \cite{py03}, dashed line,
   and AO \cite{alba}, double-dash-dotted line, with the accurate data 
 from $K_{e4}$ \cite{bnl,na48}. For AO we have not shown its error bar in order not to 
 overload the figure, see fig.1 of ref.\cite{alba} where it is given. 
\label{fig:phases}}
\end{figure}

\begin{figure}[ht]
\centerline{\epsfig{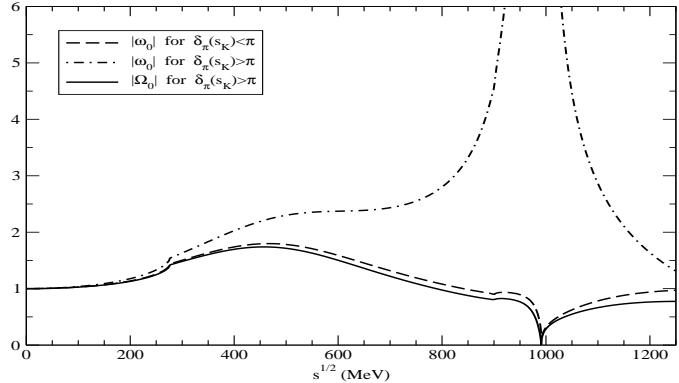}}
\vspace{0.2cm}
\caption[pilf]{\protect \small
 $|\omega_0(s)|$ from eq.(\ref{omnes})  with $\delta_\pi(s_K)_0<\pi$, dashed-line, 
 and $\delta_\pi(s_K)_0>\pi$, dot-dashed line. The solid line corresponds to the use 
  of $\om_0(s)$ in eq.(\ref{om0}) for the latter case. Here $\varphi(s)$ is used 
as  $\phi_0(s)$  in eq.(\ref{omnes}) for illustrative purposes.
\label{figomnes}}
\end{figure}

Refs.\cite{penprl,penmorgan,penanegra}   consider the Omn\`es function
$\omega_I(s)$, 
\be
\omega_I(s)=\exp\left[\frac{s}{\pi}\int_{4m_\pi^2}^\infty \frac{\phi_I(s')}
{s'(s'-s)}ds'\right]~,
\label{omnes}
\ee
with $\phi_I(s)$ the phase of $F_I(s)$ modulo $\pi$, chosen in such a way that 
 $\phi_I(s)$ is {\it continuous} and $\phi_I(4m_\pi^2)=0$. 
 This function was criticized in 
 ref.\cite{or07}. The $I=0$ S-wave  phase shift, $\delta_\pi(s)_0$, has 
a rapid increase by $\pi$ around $s_K=4m_K^2$, with $m_K$ the kaon mass,  
 due to the narrowness of the $f_0(980)$ resonance on top 
of the $K\bar{K}$ threshold. Let us denote  by $\varphi(s)$ the phase of the $\pi\pi\to\pi\pi$ 
 $I=0$ S-wave strong amplitude such that it is continuous, modulo $\pi$ when crossing a zero,
  and $\varphi(4m_\pi^2)=0$. 
 This phase is shown in fig.\ref{fig:phases} together with $ \delta_\pi(s)_0$ and $\delta_\pi(s)_2$, where the latter
 is the $\pi\pi$ $I=2$ phase shift.
   Now, if one uses $\varphi(s)$ instead of $\phi_0(s)$ in eq.(\ref{omnes})
 for illustration,  the function $\omega_0(s)$ is discontinuous 
 in the transition from $\delta_\pi(s_K)_0\to\pi-\epsilon$ to
  $\delta_\pi(s_K)_0\to \pi +\epsilon$, $ \epsilon\to 0^+$. 
 In the former case $|\omega_0(s)|$ has a zero, while in the latter it becomes $+\infty$.
 This discontinuity is illustrated in fig.\ref{figomnes} by 
 considering the difference between the  dashed and dot-dashed  lines, respectively. 
This discontinuous behaviour of $\omega_0(s)$ under small (even tiny)
 changes of $\delta_\pi(s)_0$ around $s_K$  
 was the reason for the controversy regarding the value of the pion scalar radius $\la r^2\ra^\pi_s$ 
between refs.\cite{y04,y05,y06} and ref.\cite{accgl05}. This controversy was finally solved in 
 ref.\cite{or07} where it is 
shown that Yndur\'ain's method is compatible with the solutions obtained by solving 
the Muskhelishvili-Omn\`es equations for the scalar form factor \cite{dgl,mous,moo}. 
 The problem arose because  refs.\cite{y04,y05} overlooked the proper solution 
 and stuck to an unstable one.

 Given the 
 definition of the phase function $\phi_I(s)$ in eq.(\ref{omnes}) the function 
 $F_I(s)/\omega_I(s)$ has no right hand cut. Then, refs.\cite{penprl,penmorgan,penanegra} 
 perform a twice subtracted dispersion relation of $(F_I(s)-L_I(s))/\omega_I(s)$,
 \be
F_I(s)=L_I(s)+a_I\,\omega_I(s)+c_I s\,\omega_I(s)+
\frac{s^2}{\pi}\omega_I(s)\int_{4m_\pi^2}^\infty
\frac{L_I(s')\sin\phi_I(s')}{{s'}^2(s'-s)|\omega_I(s')|}ds'~.
\label{dispen}
\ee
Low's theorem \cite{low} implies  that  $F_I\to B_I(s)$ for $s\to 0$, 
with $B_I$ the Born term contribution, shown in fig.\ref{born}.
 If we write $L_I=B_I+R_I$, with  $R_I\to 0$ for $s\to 0$, as can always be done,
 then the Low's theorem  implies also that
 $F_I-L_I\to 0$ for $s\to 0$ and hence $a_I=0$.

For the exotic $I=2$ S-wave one can invoke Watson's final state theorem\footnote{This theorem implies 
that the phase of $F_I(s)$ when there is no inelasticity is the same, modulo $\pi$,
 as the one of the isospin $I$ S-wave $\pi\pi$ elastic strong amplitude.} so that 
$\phi_2(s)=\delta_\pi(s)_2$.  
For $I=0$ the same theorem guarantees that $\phi_0(s)=\delta_\pi(s)_0$ for $s\leq 4m_K^2$. 
Here one neglects the inelasticity due to the $4\pi$ and $6\pi$
states below the two kaon threshold, an accurate assumption 
 as indicated by experiment \cite{hyams,grayer}.
 Above the two kaon threshold $s_K=4m_K^2$, 
  the phase function $\phi_0(s)$ cannot 
be fixed {\it a priori} due 
the onset of inelasticity. This is why ref.\cite{penprl} took  for $s>s_K$ that either
$\phi_0(s)\simeq \delta_\pi(s)_0$ or $\phi_0(s)\simeq \delta_\pi(s)_0-\pi$, in order
 to study the size of the  uncertainty induced for lower energies. It results, as mentioned above, 
 that  this uncertainty becomes huge already for $\gtrsim 0.55$~GeV \cite{penprl}.

 Let us now explain how this problem was settled in ref.\cite{orsletter}. 
Inelasticity is again small for $1.1 \lesssim \sqrt{s}\lesssim 1.5$~GeV  \cite{hyams,grayer}. 
 As remarked in refs.\cite{y04,or07}, one can then apply 
 approximately Watson's final state 
 theorem and for $F_0(s)$ this implies that 
  $\phi_0(s)\simeq \delta^{(+)}(s)$ modulo $\pi$. Here  $\delta^{(+)}(s)$ is the eigenphase 
  of the $\pi\pi$, $K\bar{K}$ $I=0$ S-wave S-matrix 
 such that it is continuous and $\delta^{(+)}(s_K)=\delta_\pi(s_K)_0$. In refs.\cite{y05,or07} it is
 shown that $\delta^{(+)}(s)\simeq \delta_\pi(s)_0$ or $\delta_\pi(s)_0-\pi$, depending on whether 
 $\delta_\pi(s_K)_0\geq \pi$ or $<\pi$, respectively. In order to fix the 
integer factor in front of $\pi$ in the relation 
$\phi_0(s)\simeq \delta^{(+)}(s)$ modulo $\pi$, one needs to devise an argument to follow the possible 
 trajectories of  $\phi_0(s)$ 
 in the {\it narrow} region $1\lesssim \sqrt{s}\lesssim 1.1$~GeV,  
where inelasticity is not negligible. The remarkable physical effects happening there 
 are the appearance of the 
 $f_0(980)$ resonance on top of the $K\bar{K}$ threshold and the cusp effect of the latter 
 that induces a discontinuity  at $s_K$ in the derivative  of observables, this is clearly visible 
 in  fig.\ref{fig:phases}. 
 Between 1.05 to 1.1~GeV there are no further narrow structures and observables evolve smoothly. 
 Approximately half of the region between 
0.95 and 1.05~GeV is elastic and $\phi_0(s)= \delta_\pi(s)_0$ (Watson's theorem), so that it raises 
 rapidly. Above $2 m_K\simeq 1$~GeV and up to 1.05~GeV the function $\phi_0(s)$ can keep increasing with 
 energy, like $\delta_\pi(s)_0$ or $\varphi(s)$ for $\delta_\pi(s_K)_0\geq \pi$, and
this is always the case for the corresponding phase function  of
 the strange scalar form factor of the pion \cite{or07}. It is also the behaviour 
 for $\phi_0(s)$ corresponding to the explicit calculation of ref.\cite{oo98}.  
 The other possibility is a change  of sign in the slope at $s_K$ due to the 
$K\bar{K}$ cusp effect such 
 that $\phi_0(s)$  starts a rapid decrease in 
 energy, like $\varphi(s)$ for $\delta_\pi(s_K)_0<\pi$, fig.\ref{fig:phases}.
  Above $\sqrt{s}=1.05$~GeV,  $\phi_0(s)$ matches smoothly with the 
 behaviour for $\sqrt{s}\gtrsim 1.1$~GeV, which it is worth recalling that  it is constraint 
 by Watson's final state theorem. 
   As a result of this matching, for $\sqrt{s}\gtrsim 1$~GeV 
  {\it either}  $\phi_0(s)\simeq \delta_\pi(s)_0$ {\it or}
   $\phi_0(s)\simeq \delta_\pi(s)-\pi$, corresponding
 to an increasing or decreasing $\phi_0(s)$  above $s_K$, in order.  
 Our argument also justifies the similar choice of phases in ref.\cite{penprl} above $s_K$ to estimate
 uncertainties.   
 
 Let us define the switch $z$ to characterize the behaviour of $\phi_0(s)$
  for $s>s_K$, and close to $s_K$,   such that
  $z=+1$ if $\phi_0(s)$ rises with energy and $z=-1$ if it decreases.
 Let $s_1$ be the value of $s$ at which $\phi_0(s_1)=\pi$. 
 As shown in ref.\cite{or07}, instead of eq.(\ref{omnes}), one should better consider the 
  Omn\`es function,
\be
\om_0(s)=\left(1-\theta(z)\frac{s}{s_1}\right)
\exp\left[ \frac{s}{\pi}\int_{4m_\pi^2}^\infty \frac{\phi_0(s')}{s'(s'-s)} ds'\right]~,
\label{om0}
\ee
with $\theta(z)=1$ for $z=+1$ and 0 for $z=-1$. 
 Because of the first order polynomial in $\om_0(s)$, with a zero at $s_1$, one can guarantee a 
 continuous behaviour for $\om_0(s)$ when moving from $z=-1$ to $z=+1$, passing  
 from a deep minimum  to a zero. This is shown in fig.\ref{figomnes}  by comparing the solid 
line with the dashed one, both computed with eq.(\ref{om0}) and using $\varphi(s)$ as $ \phi_0(s)$ for illustrative
purposes. Both 
lines are very close to each other as corresponds to a continuous transition. 
 Note that $s_1$ is the only point around $s_K$ where 
the imaginary part of $\omega_0(s)$ vanishes, and this fixes unambiguously the position 
of the zero in eq.(\ref{om0}), as discussed in ref.\cite{or07}.   
For $I=2$ since the phase shifts are small and smooth \cite{losty,hoog}, see fig.\ref{fig:phases}, 
the issue of the discontinuity  under changes in the parameterizations 
does not arise and we keep using $\omega_2(s)$, eq.(\ref{omnes}).

Now, we perform a twice subtracted dispersion relation for $(F_0(s)-L_0(s))/\om_0(s)$, 
employing $\om_0(s)$ instead of $\omega_0(s)$. Instead of eq.(\ref{dispen}) one obtains
\be
F_0(s)=L_0(s)+c_0 s\om_0(s)+
\frac{s^2}{\pi}\om_0(s)\int_{4m_\pi^2}^\infty
\frac{L_0(s')\sin\bar\phi_0(s')}{{s'}^2(s'-s)|\om_0(s')|}ds'
+\theta(z)\frac{\omega_0(s)}{\omega_0(s_1)}\frac{s^2}{s_1^2}(F_0(s_1)-L_0(s_1))~.
\label{f0f}
\ee

In the previous equation we introduce $\bar\phi_0(s)$ that is defined 
 as the phase of $\om_0(s)$.  Let us note that in the case 
 $z=+1$  the phase of $\om_0(s)$ for $s>s_1$ 
 is not $\phi_0(s)$ but $\phi_0(s)-\pi$, 
 due to the factor $1-(s+i\epsilon)/s_1$ in $\Omega_0(s)$, eq.(\ref{om0}),
 present in this case. 
 For $I=2$  we  use eq.(\ref{dispen}). 

It is worth mentioning 
 that eq.(\ref{f0f}) for $I=0$ and $z=+1$ is equivalent to perform a three times subtracted
dispersion relation for $(F_0(s)-L_0(s))/\omega_0(s)$,
 instead of the twice subtracted dispersion relation in eq.(\ref{dispen}) as proposed in 
 refs.\cite{penprl,penmorgan,penanegra}. 
 In eq.(\ref{f0f}), two subtractions are taken at $s=0$ and a third one 
 at $s_1$. We could have taken the third subtraction at $s=0$ as well, so that for
 $z=+1$ one has
 \be
F_0(s)=L_0(s)+c_0 s\, w_0(s)+d_0 s^2 w_0(s)+\frac{s^3 w_0(s)}{\pi}\int_{4m_\pi^2}^\infty \frac{L_0(s')
\sin\phi_0(s')}{s'^3(s'-s)|\omega_0(s')|}ds'~,
\label{f0fthree}
 \ee
  Nevertheless, we 
 find more convenient eq.(\ref{f0f}) as it is physically motivated by the use
  of the Omn\`es function eq.(\ref{om0}) that is continuous
   under changes in the parameterization of the $I=0$ 
  S-wave S-matrix. Furthermore, eq.(\ref{f0f}) is more suited to constraint the $f_0(980)$ 
  energy region in order to implement the condition {\bf 3} below, because $s_1\simeq 1$~GeV$^2$ and 
  it coincides very approximately with the mass of this resonance.

Let us denote by $F_N(s)$ the  S-wave $\gamma\gamma\to \pi^0\pi^0$ amplitude and by $F_C(s)$ the
$\gamma\gamma\to\pi^+\pi^-$ one. 
The relation between $F_0$, $F_2$  and $F_N(s)$, $F_C(s)$ is 
\ba
F_N(s) &=& -\frac{1}{\sqrt3} F_0 + \sqrt{\frac23} F_2~, \nn \\
F_C(s) &=& -\frac{1}{\sqrt3} F_0 - \sqrt{\frac16} F_2 ~.
\ea

We are still left with the unknown constants $c_0$, $c_2$ for $I=0$ and $2$, respectively, and 
$F_0(s_1)-L_0(s_1)$ for $I=0$ and $z=+1$.  In order to determine them we impose the 
following conditions: 

\noindent
{\bf 1.} $F_C(s)-B_C(s)$ vanishes linearly in $s$ for $s\to 0$ and we match the
 coefficient to the one loop $\chi$PT result \cite{bc88,dhl87}. 

\noindent  
{\bf 2.} $F_N(s)$  vanishes linearly for $s\to 0$ as well and the coefficient can be 
 obtained again from  one loop $\chi$PT \cite{bc88,dhl87}.

\noindent
{\bf 3.} For $I=0$ and $z=+1$ one has in addition the constant $F_0(s_1)-L_0(s_1)$. Its value 
 can be restricted taking into account that $F_N(s)$ has an Adler zero, 
due to chiral symmetry. This zero is located at $s_A= m_\pi^2$ in one loop $\chi$PT and 
 moves to $s_A=1.175 m_\pi^2$ in two loop $\chi$PT \cite{bgs94}. A rather large
 correction around 18$\xpc$ results, which prevents us from taking $s_A$ straightforwardly 
 as given by $\chi$PT. In turn, we obtain that the value of the resulting cross section
  $\sigma(\gamma\gamma\to\pi^0\pi^0)$ around the $f_0(980)$ resonance is quite sensitive
  to the value of $F_0(s_1)-L_0(s_1)$. This is so because this constant
  appears in the last term in eq.(\ref{f0f}) which dominates 
 the energy region around the $f_0(980)$ since $\Omega_0(s_1)=0$. 
 Though the  dispersive method is devised at its best 
for lower energies, it is also clear that it should give, at least, the proper 
order of magnitude for $\sigma(\gamma\gamma\to \pi^0\pi^0)$ in
  the $f_0(980)$ region. E.g., other models \cite{oo98,acharev,mennessier} having similar 
 basic ingredients describe that energy region quite well indeed. 
  We then restrict  $F_0(s_1)-L_0(s_1)$ 
 so that $\sigma(\gamma\gamma\to \pi^0\pi^0)\leq 40$~nb at $s_1$. Note that 
the D-wave  component of this cross-section at $s_1\simeq 1~$GeV, due to the tail of 
the $f_2(1270)$ resonance, is not longer small and hence the S-wave contribution should be 
significantly smaller than the experimental value $<40$~nb. 
 This is clearly shown by experiment \cite{crystal}, 
  see fig.\ref{fig:fignew2}. The S- and D-wave  contributions add incoherently 
in the total cross section because of the integration  on  the azimuthal angle as the $f_2(1270)$
resonance signal is overwhelming helicity 2 \cite{penmorgan}.
 We have  checked that this upper bound for the peak of the $f_0(980)$ 
in $\gamma\gamma\to\pi^0\pi^0$ is equivalent to impose that the $\gamma\gamma$ width of the 
$f_0(980)$, $\Gamma(f_0(980)\to\gamma\gamma)$, lies in the range
$205^{+95}_{-83}(stat)^{+147}_{-117}(sys)$~eV, determined by the Belle Collaboration 
in ref.\cite{mori}. 
The width $\Gamma(f_0(980)\to \gamma\gamma)$ is calculated analogously as done in section
 \ref{sec:sigwidth} for 
the $\gamma\gamma$ width of the $\sigma$ resonance. We shall see that the effect of this rather large uncertainty
 allowed at 1~GeV, see fig.\ref{fig:fignew2}, is very mild 
at lower energies, as shown in fig.\ref{fig:updown}.

It is worth discussing further the improvement achieved by eq.(\ref{f0f}), first derived in 
ref.\cite{orsletter}, in the understanding of 
the role played by  the $f_0(980)$ resonance in $\gamma\gamma\to \pi\pi$. 
The resulting $|F_0(s)|$ given by eq.(\ref{dispen}) from ref.\cite{penprl} explodes for $z=+1$ 
 around $s_1$, as $|w_0(s)|$ becomes extremely large for this energy as shown in fig.\ref{figomnes} by the 
dot-dashed line.
  To avoid this large number the coefficient of $w_0(s_1)$ in eq.(\ref{dispen}) is required to almost vanish, 
\be
c_0 +\frac{s_1}{\pi}\int_{4m_\pi^2}^\infty \frac{L_0(s') \sin\phi_0(s')}{{s'}^2(s'-s_1)|\omega_0(s')|}\simeq
0~.
\label{finetun}
\ee
Thus,  a dramatic fine tuning of $\phi_0(s)$ is implied,
 so as the integral above must almost cancel
 $c_0$. Note that $c_0$ is a real number and 
  this integral is  real only at $s=s_1$, precisely where it is evaluated. Then the cancellation 
 in eq.(\ref{finetun}) is possible. 
 But $c_0$, following the approach of ref.\cite{penprl,penanegra}, is already determined
 by the requirement that the Adler zero for $F_N(s)$ is at $s_A=m_\pi^2/2$, $m_\pi^2$ 
 or $2 m_\pi^2$ \cite{penprl}. The subsequent uncertainty in $c_0$, together with the uncertainty in the  
 knowledge of  $\phi_0(s)$  above the $K\bar{K}$ threshold, invalidates in practice the fine tuning 
 in eq.(\ref{finetun}). This makes that the results of ref.\cite{penprl} 
 for $\sigma(\gamma\gamma\to \pi^0\pi^0)$ diverge rapidly with energy 
above $\sqrt{s}\simeq 0.5$~GeV for $z=+1$ as remarked above,
 see fig.3 of ref.\cite{penprl}. Refs.\cite{penanegra,penmorgan,dh93} did not discuss 
 this source of large uncertainty in
their calculations but, since these papers follow the same scheme as ref.\cite{penprl}, it is 
 certainly present. 
One of the main results that follow from eq.(\ref{f0f})  
is to showing that by using $\om_0(s)$ 
 one can cure such diverging results for 
 $z=+1$.  This issue is solved 
 because for $s$  around $s_1$ one  has that $\om_0(s)\simeq 0$ and the diverging effects 
due to $\omega_0(s)$ for $z=+1$ are isolated in the last term and controlled 
 by the factor $F_0(s_1)-L_0(s_1)$. The small signal of the $f_0(980)$ implicit in 
the $\sigma(\gamma\gamma\to \pi^0\pi^0)$ data of ref.\cite{crystal}, and explicitly seen in 
$\gamma\gamma\to \pi^+\pi^-$ \cite{mori}, implies a vanishing value of the constant $F_0(s_1)-L_0(s_1)$.
In our actual calculations we obtain typically $(F_0(s_1)-L_0(s_1))/L_0(s_1)\lesssim 10^{-2}$.
 One can indeed understand the smallness of this number 
 by a continuity argument. 
For that, let us envisage a mathematical continuous transition 
from $z=-1$ to $z=+1$ by taking $\varphi(s)$ as $\phi_0(s)$ in eq.(\ref{f0f}). The situation $z=-1$ is
accomplished when $\delta_\pi(s_K)_0<\pi$ and $z=+1$ when $\delta_\pi(s_K)_0\geq \pi$, see 
fig.\ref{fig:phases}. Now, imposing that $F_0(s)$ is continuous under the 
 transition $\delta_\pi(s_K)_0\to \pi-\epsilon$ (where the last term in eq.(\ref{f0f}) is absent)
  to $\delta_\pi(s_K)_0\to \pi+\epsilon$,  $\epsilon\to 0^+$, where it is present, implies that 
  $F_0(s_1)-L_0(s_1)=0$ for $\delta_\pi(s_K)_0=0$. Since $\phi_0(s)$ is quite close to $\varphi(s)$ 
both  for 
 $z=+1$ and $-1$, the continuous transition considered is not just academic because 
 $\delta_\pi(s_K)\simeq \pi$.\footnote{See ref.\cite{y05} for a detailed discussion on the value 
 of $\delta_\pi(s_K)$.}   As a result one should expect 
 that $F_0(s_1)-L_0(s_1)\simeq 0$.

 Since the $f_0(980)$ resonance gives rise to a small {\it peak} in the precise data 
 on $\gamma\gamma\to\pi^+\pi^-$ \cite{mori}, then $\phi_0(s)$ must increase with energy above $s_K$
and the case with $z=+1$ is the one realized in nature. As a result
 $\phi_0(s)\simeq \delta_\pi(s)_0$, at least up to 
around $\sqrt{s}\lesssim 1.5$~GeV. Otherwise $z=-1$ in eq.(\ref{f0f}) and because $|\omega_0(s)|$ has a deep 
around the $f_0(980)$ mass there is no a local maximum associated with this resonance in $|F_0(s)|$ but a minimum. 
 One should keep in mind  that  $z=+1$  and 
 only this case will be considered in the subsequent.

In ref.\cite{orsletter} both cases $z=\pm 1$  were discussed and kept, so that the differences 
in the results by taking one case or the other were included in the uncertainties 
of the approach, e.g.  when presenting  $\sigma(\gamma\gamma\to \pi^0\pi^0)$ 
for $\sqrt{s}\lesssim 0.8$~GeV.
 In the latter reference  the main focus was driven
towards the drastic reduction of the uncertainty in the results
 for the two choices of $\phi_0(s)$ above $s_K$.  Here we are 
more accurate as we have singlet out the $z=+1$ case as the proper one for $\phi_0(s)$  and 
$s>s_K$. 
 
We can estimate the uncertainty in the approximate relation $\phi_0(s)\simeq \delta_\pi(s)_0$ 
for  $4m_K^2\lesssim \sqrt{s}\lesssim 1.5$~GeV by similar arguments to those employed in ref.\cite{or07}. 
Let the  $\pi\pi$ and $K\bar{K}$ $I=0$ S-wave  $S-$matrix, 
\ba
S=\left(
\begin{array}{cc}
\eta e^{2i\delta_\pi} & i\sqrt{1-\eta^2}e^{i(\delta_\pi+\delta_K}) \\
i\sqrt{1-\eta^2}e^{i(\delta_\pi+\delta_K}) & \eta e^{2i\delta_K}
\end{array}
\right)~,
\label{smat}
\ea
with $\eta$ the elasticity parameter. This matrix is diagonalized 
 with the orthogonal matrix $C$,
\be
C=\left(
\begin{array}{cc}
\cos\theta & \sin \theta \\
-\sin \theta & \cos \theta
\end{array}
\right)~.
\label{cmat}
\ee
In eq.(\ref{smat}) $\delta_\pi\equiv \delta_\pi(s)_0$ and $\delta_K$ is the kaon phase shifts. 
 Explicit expressions for $\cos\theta$ and $\sin\theta$ in terms of the $I=0$
 S-wave $\pi\pi$ and $K\bar{K}$ phase shifts 
and the elasticity parameter are given in ref.\cite{or07}.  In this reference one can also find the
 expression for the eigenvalues $e^{2i\delta^{(+)}}$ and $e^{2i\delta_{(-)}}$.  
This diagonalization allows to disentangle two elastic scattering channels and then the two photon 
$I=0$ S-wave  amplitudes attached to every of these channels, $\Gamma_1'$ and $\Gamma_2'$,
 will satisfy the Watson's final state theorem in the whole
energy range. Because of this diagonalization one has,
\ba
F&\equiv& \left(
\begin{array}{c}
F_0\\
F_K
\end{array}\right)
=C F'=C \left(
\begin{array}{c}
\Gamma_1' \\
\Gamma_2'
\end{array}\right)~,\nn\\
F_0 &=& \cos \theta \,\Gamma_1' +  \sin \theta \, \Gamma_2'~,\nn\\
F_K &=&   \cos \theta \, \Gamma_2' -\sin \theta\, \Gamma_1'~.
\label{diag}
\ea
Here $F_K$ is the  $I=0$ S-wave $\gamma\gamma\to K\bar{K}$ amplitude. 
Now, when $\eta\to 1$ then $\sin\theta\to 0$ as $\epsilon=\sqrt{(1-\eta)/2}$ \cite{or07} and
$F_0(s)$  is given approximately by $\Gamma_1'(s)$. With 
$\tan\theta \,\Gamma'_2/\Gamma'_1=\epsilon \,e^{i\rho}$, we rewrite $F_0$ above as,
\ba
F_0(s)&=&\cos\theta \,\Gamma'_1\left[1+\epsilon (\cos\rho+i\sin\rho) \right]\nn\\
&=&\cos\theta \, \Gamma'_1 (1+\epsilon \cos\rho)\left(1
+\frac{i\epsilon\sin\rho}{1+\epsilon\cos\rho}\right)\nn\\
&=&\cos\theta \, \Gamma'_1 (1+\epsilon \cos\rho)e^{i\lambda}+{\cal O}(\epsilon^2)~.
\ea
with $\lambda=\epsilon\,\sin\rho/(1+\epsilon \cos \rho)$. From the previous equation it follows 
that 
\be 
\phi_0(s)=\delta_\pi(s)_0+\lambda+{\cal O}(\epsilon^2)
\label{correction}
\ee with $\epsilon\to 0$ for $\eta\to 1$. 
 Using $\eta=0.8$ in eq.(\ref{correction}) for the range 
$1.1\lesssim \sqrt{s}\lesssim 1.5$~GeV (from the K-matrix in eq.(\ref{km}) $\eta\simeq 1$), one ends 
with $|\epsilon|\simeq 0.3$.  Taking into account that $\delta_\pi(s)_0$ is larger than $3\pi/2$, 
 fig.\ref{fig:phases}, then one has a relative correction of around a $10\xpc$. 
 We can also proceed similarly in the region $1 \lesssim \sqrt{s} \lesssim 1.1$~GeV. Applying 
 now eq.(\ref{correction}) with 
 $\eta\gtrsim 0.6$ from the $\pi\pi\to K\bar{K}$ scattering experiments \cite{1920or07}, and taking 
 $\delta_\pi(s)_0\geq \pi$ in the previous energy range, we end up with a $\sim 20\xpc$ uncertainty.
  We shall incorporate in our error analysis 
the previous uncertainties.

 We now consider the phases used above $s_H\equiv 2.25$~GeV$^2$. For $I=2$, where final state interaction
 give rise to small corrections, we directly take the extrapolation of our fits. Notice that due 
 to the way that the vector and axial-vector exchanges are calculated from the chiral Lagrangians 
 presented in the next section, one has that $L_I(s)$ diverges linearly in $s$ for $s\to +\infty$. On the
 other hand $|\omega_I(s)|\propto s^{-\nu_I}$ with $\nu_I=\phi_I(\infty)/\pi$. 
Taking this into account for $I=2$ the
 integrand in eq.(\ref{dispen}) tends to $\sin\phi_2(s') {s'}^{\nu_2-1}/(s'-s)$ for $s'\to+\infty$,
  with $|\nu_2|<<1$,  and then the dispersion integral  converges. 
 For $I=0$ we have that $|\Omega_0(s)|$ behaves as $s^{-\nu_0+1}/s_1$ for $s\to +\infty$. 
 We know already $\phi_0(s)$ for $s$ quite high,
   and noticing that $\phi_0(s_H)/\pi\simeq 2$ we take this value as a good approximation 
   for $\nu_0$ so that 
$\Omega_0(s)\propto s^{-1}$ and $\phi_0(+\infty)=2\pi$. The integrand in eq.(\ref{f0f}) tends then to 
$\sin\phi_0(s')/(s'-s)$ and converges because $\sin\phi_0(\infty)\to 0$. As a soft extrapolation for
$\phi_0(s)$ for $s>s_H$ we shall take the same expression as in ref.\cite{or07} with $2\pi$ as
 the limiting value,
\be
\phi_0(s)=2\pi\left(1\pm\frac{1}{\log s/\Lambda^2}\right)~,~s>s_H~,
\label{asintotico}
\ee
with $\Lambda^2\in[0,0.35]$~GeV$^2$. In this way at  $s=s_H$ one has 
an uncertainty of $\sim \pm \pi/2$ degrees that narrows only
logarithmically for higher energies. This uncertainty is incorporated in the error analyses that
follow. We think that our choices for $\phi_I(s)$ for $s>s_H$, and the  uncertainties taken,  
 are quite reasonable. Let us note that for $s<1$~GeV$^2$ 
 the integrand in the dispersive integral is overweightly dominated for 
$s'<<s_H$.

\section{Dynamical approach to $\gamma\gamma\to \pi\pi$. }
\label{sec:dyn}

In this section we review and update the approach of ref.\cite{oo98} based on
U$\chi$PT. This approach  extends the $\chi$PT one loop calculation
of $\gamma\gamma\to \pi^0\pi^0$, $\pi^+\pi^-$ and $K^+K^-$,
 implementing unitarity in coupled channels, in order  
to account of the meson-meson final state interactions, and by including vector and axial-vector 
resonance exchanges.   For the derivation of the unitarization formulae that we use below 
we refer to the original  ref.\cite{oo98}, see also refs.\cite{sfm,vfm} for later advances in the
formalism.
 In the dispersive method discussed in the
previous section this is accomplished by the use of the Omn\`es functions $\Omega_0(s)$ and 
$\omega_2(s)$ in the  dispersion relations eqs.(\ref{f0f}) and (\ref{dispen}), respectively.

On the other hand, we also offer here the expressions for the Born terms and the crossed
exchanges of vector ($1^{--}$) and axial-vector ($1^{++}$, $1^{+-})$ resonances, 
employed both in the dispersive and dynamical approaches.
 The exchange of vector resonances calculated from chiral Lagrangians
was  performed in ref.\cite{chino}, while the inclusion of the
 $1^{++}$ axial ones was undertaken in ref.\cite{chino2}. Here we have 
calculated these tree level graphs but our results are different to those of ref.\cite{chino2}
 for the latter diagrams. Reassuringly, we agree with the expressions given in ref.\cite{bgs94} for the 
 $1^{+-}$ axial-vector resonance exchanges, which are also included here, 
 and that share the same Lorentz structure as the $1^{++}$ resonance exchange contributions. 
 It is worth mentioning that 
the $1^{++}$  resonance exchanges contribute already 
at ${\cal O}(p^4)$ in $\chi$PT for $\gamma\gamma\to \pi^+\pi^-$, 
while the other resonance exchanges start one order higher in $\gamma\gamma\to\pi\pi$. 
The formulae corresponding to the Born term
and the different resonance exchanges are  given below.

\subsection{Formulae for $F_N$ and $F_C$}
The expressions for $F_N(s)$ and $F_C(s)$ of ref.\cite{oo98} are enlarged now to 
include the exchanges of the $K^*$ and the $1^{+-}$ axial-vector nonet: 
\ba
F_N&=&T_{\omega,\pi^0\pi^0}^S+T_{C,\pi^0\pi^0}^S+T_{\rho,\pi^0\pi^0}^S+\left(\widetilde{t}_{\rho\pi^0\pi^0}
+\widetilde{t}_{\omega\pi^0\pi^0}+\widetilde{t}_{C\pi^0\pi^0}\right)t_{\pi^0\pi^0,\pi^0\pi^0}\nn\\
&&+\left(\widetilde{t}_{\chi \pi}+\widetilde{t}_{\rho\pi^+\pi^-}+\widetilde{t}_{A\pi^+\pi^-}
+\widetilde{t}_{C\pi^+\pi^-}\right)t_{\pi^+\pi^-,\pi^0\pi^0}\nn\\
&& +\left(\widetilde{t}_{\chi K}+2\widetilde{t}_{K^*K^+K^-}+\widetilde{t}_{AK^+K^-}+
5 \widetilde{t}_{CK^+K^-}\right)t_{K^+K^-,\pi^0\pi^0}~,\nn\\
F_C&=& T^S_B +  T_{\rho,\pi^+\pi^-}^S+ t^S_{A,\pi^+\pi^-} +  t^S_{C,\pi^+\pi^-} 
+\left(\widetilde{t}_{\rho\pi^0\pi^0}
+\widetilde{t}_{\omega\pi^0\pi^0}+\widetilde{t}_{C\pi^0\pi^0}\right)t_{\pi^0\pi^0,\pi^+\pi^-}\nn\\
&&+\left(\widetilde{t}_{\chi \pi}+\widetilde{t}_{\rho\pi^+\pi^-}+\widetilde{t}_{A\pi^+\pi^-}
+\widetilde{t}_{C\pi^+\pi^-}\right)t_{\pi^+\pi^-,\pi^+\pi^-}\nn\\
&& + \left(\widetilde{t}_{\chi K}+2\widetilde{t}_{K^*K^+K^-}
+\widetilde{t}_{AK^+K^-}+5\widetilde{t}_{CK^+K^-}\right)t_{K^+K^-,\pi^+\pi^-}~.
\label{fcnpa}
\ea
The different terms in this equation are given below. 
Although not explicitly indicated in the subscripts, the contributions from amplitudes 
with the intermediate  $K^0\bar{K}^0$ state 
 are encoded in the brackets proportional to $t_{K^+K^-,\pi\pi}$, because 
this S-wave strong amplitude is the same as $t_{K^0\bar{K}^0,\pi\pi}$ 
and kaon masses are all taken the same (isospin limit). 

In the previous equations the strong S-wave amplitudes can be obtained from those 
 with isospin defined as,
\ba
t_{\pi^+\pi^-,\pi^+\pi^-}&=&\frac{2}{3}t^{I=0}_{\pi\pi,\pi\pi}+\frac{1}{3}t^{I=2}_{\pi\pi,\pi\pi}~,\nn\\
t_{\pi^0\pi^0,\pi^0\pi^0}&=&\frac{2}{3}t^{I=0}_{\pi\pi,\pi\pi}+\frac{4}{3}t^{I=2}_{\pi\pi,\pi\pi}~,\nn\\
t_{\pi^+\pi^-,\pi^0\pi^0}&=&\frac{2}{3}t^{I=0}_{\pi\pi,\pi\pi}-\frac{2}{3}t^{I=2}_{\pi\pi,\pi\pi}~,\nn\\
t_{K^+K^-,\pi^0\pi^0}&=&t_{K^+K^-,\pi^+\pi^-}=\frac{1}{\sqrt{3}}t^{I=0}_{K\bar{K},\pi\pi}~.
\ea
In ref.\cite{oo98} the $I=0$ and $I=2$ S-wave amplitudes were calculated as in ref.\cite{npa}. 
 For $I=2$ we employ directly 
the experimental phase shifts   
 $t_{\pi\pi,\pi\pi}^{I=2}=-\frac{8\pi\sqrt{s}}{q_\pi}e^{i\delta_\pi(s)_2}\sin \delta_\pi(s)_2$.

In eq.(\ref{fcnpa}) the amplitudes $T_{\omega,\pi^0\pi^0}$, $T_{\rho,\pi\pi}$, 
$T_{K^*,K \bar{K}}$  are the crossed exchange of
the $\omega$, $\rho$ and $K^*$ vector resonances, in order, denoted generically by 
$T_{V,P\bar{P}}$. On the other hand, $T_{A,P\bar{P}}$ and 
$T_{C,P\bar{P}}$ refer to the exchange of the $1^{++}$ and $1^{+-}$ multiplets of 
axial-vector resonances, respectively.  
In addition, $\widetilde{t}_{\chi P}$ represents the one loop $\chi$PT function,
\be
\widetilde{t}_{\chi P}(s)=\frac{2e^2}{16\pi^2}\left\{1+
\frac{m_P^2}{s}\left[\log\frac{1+\sigma_P(s)}{1-\sigma_P(s)}
-i\pi\right]^2\right\}~,
\ee
with $m_P$ the mass of the pseudoscalar $P$, either $\pi$ or $K$, 
and $\sigma_P(s)=\sqrt{1-4m_P^2/s}$. Finally, $\widetilde{t}_{RP\bar{P}}$ corresponds
 to the unitarity loop with the pseudoscalars $P$,
$\bar{P}$ and the crossed exchange of the resonance $R$, either vector or axial, calculated 
as in  ref.\cite{oo98}.

Here we give the expressions calculated for the Born terms, fig.\ref{born}, 
$\gamma(k_1,\lambda_1)\gamma(k_2,\lambda_2)\to P^+(p_1)P^-(p_2)$,
\be
T_B=2e^2\left[\epsilon_1\epsilon_2-\frac{(\epsilon_1 p_1)(\epsilon_2 p_2)}{p_1 k_1-i0^+}
-\frac{(\epsilon_1 p_2)(\epsilon_2 p_1)}{p_1 k_2 -i0^+}\right]~,
\ee
where $\epsilon_1(k_1,\lambda_1)$, $\epsilon_2(k_2,\lambda_2)$ 
are the polarization vectors of the photons taken in the
gauge $\epsilon_1 k_1=\epsilon_1 k_2=\epsilon_2 k_1=\epsilon_2 k_2=0$, with 
polarizations given by $\lambda_1$ and $\lambda_2$, respectively. 
 In the following, we shall only need the three-momenta $\vec{k}_1$ and $\vec{k}_2$
  along the $z$-axis and 
use the four-vectors $(0,\vec{\epsilon}(k,\lambda))$ 
 with  $\vec{\epsilon}(k\widehat{z},\pm)=\mp \frac{1}{\sqrt{2}}(\widehat{x}\pm i
 \widehat{y})$ and $\vec{\epsilon}(-k\widehat{z},\pm)=\mp \frac{1}{\sqrt{2}}(\widehat{y}\pm i
 \widehat{x})$. The projection of the Born term in S-wave is given by
 $i \int d\widehat{p}\; T_B(\lambda_1=+,\lambda_2=+,\vec{p})/4\pi$, with $\vec{p}$ the 
 CM three-momentum of the pions. The prefactor $i$ is included
 for convenience. An analogous formula for the projection in S-wave is employed for any other
 amplitude of $\gamma\gamma \to P\bar{P}$. We then have,
\be
T_B^S= 2 e^2 \frac{2 m_\pi^2/s}{\sigma_\pi(s)}\log\frac{1+\sigma_\pi(s)}{1-\sigma_\pi(s)}~,
\label{tbs}
\ee
with $\sigma_\pi=\sqrt{1-4m_\pi^2/s}$.

We now consider the crossed exchange of the vector resonances in 
$\gamma(k_1,\lambda_1)\gamma(k_2,\lambda_2)\to P(p_1)\bar{P}(p_2)$.
 We evaluate them employing the chiral Lagrangian 
\cite{chino},
\be
{\cal L}_{V P\gamma}=R_\omega e \,\epsilon_{\mu \nu \alpha \beta} F^{\mu \nu}\hbox{Tr}\left[ \Phi
\left\{Q,\partial^\alpha V^\beta\right\}\right]~,
\ee
where the matrix fields $\Phi(x)$, $V(x)$ contain the octet of pseudoscalars and vectors \cite{chino} 
and $Q$ is the
quark charge matrix, $Q=diag(2/3,-1/3,-1/3)$.
\be
\begin{aligned}
T_{V,P\bar{P}}=e^2\,R^2_V& \left[\frac{(\epsilon_1 \epsilon_2)[s(t+m_P^2)-(t-m_\pi^2)(u-m_\pi^2)]
-2s(\epsilon_1 p_1)(\epsilon_2 p_1)}{t-M_V^2+i0^+}\right.\\
&\left.+\frac{(\epsilon_1 \epsilon_2)[s(u+m_P^2)-(t-m_\pi^2)(u-m_\pi^2)]-2s(\epsilon_1 p_1)
(\epsilon_2 p_1)}{u-M_V^2+i0^+}\right]
\label{expv}
\end{aligned}
\ee
with $P=\pi^0,$ $\pi^+$, $K^+$ or $K^0$and $V=\omega$, $\rho$ or $K^*$. Furthermore,
$R_\rho^2=R^2_{K^*}=R^2_\omega/9$. In these expressions $M_\omega$, $M_\rho$ and $M_{K^*}$
are the masses of the $\omega$, $\rho$ and $K^*$, respectively. The coupling $R^2_\omega$
is determined from the width $\Gamma(\omega\to \gamma \pi^0)$,
\be
R_\omega^2=\frac{6}{\alpha}\frac{M_\omega^3}{(M_\omega^2-m_\pi^2)^3}
 \Gamma(\omega\to \gamma \pi^0)=1.44~\hbox{GeV}^{-2}.
\ee

The projection of eq.(\ref{expv}) is given by,
\be
T_{V,P\bar{P}}^S=-2e^2\,R^2_V\left[\frac{M_V^2}{\sigma_P}\log
\frac{1+\sigma_P+s_V/s}{1-\sigma_P+s_V/s}-s\right]~,
\label{expvs}
\ee
with  $s_V=2(M_V-m_P^2)$.
It is important to remark that  eqs.(\ref{expv}) and (\ref{expvs}) 
contribute at the level of a two loop $\chi$PT calculation in $\gamma\gamma\to\pi\pi$.
 This implies that vector exchange
is relatively suppressed at low energies  due to its rather high chiral counting. Nonetheless 
in fig.\ref{fig:BVA} it is shown that they are significant above 0.4~GeV.
 The $1^{++}$ axial-vector crossed exchanges, as we shall see immediately,
 appear already at the level of one loop $\chi$PT. Hence, for low energy
$\gamma\gamma\to\pi\pi$ their contributions
are more important than those due to the exchange of vectors.  
 To evaluate the contribution from the crossed exchanges of the $1^{++}$ 
axial resonances in $\gamma\gamma\to
\pi^+\pi^-$ and $\gamma\gamma\to K^+K^-$ we consider the Lagrangian \cite{colla},
\be
{\cal L}_{A P\gamma}=i\frac{e F_A}{2f_\pi}F^{\mu \nu}\hbox{Tr}\left(\Phi[Q,A_{\mu\nu}]\right)~,
\ee
with $A_{\mu\nu}$ a matrix with the $1^{++}$ octet of axial resonances and $f_\pi$ the pion decay constant.
 Here the antisymmetric tensor formalism for the axial-vector field is employed. 
 The resulting amplitudes are
\ba
\begin{aligned}
T_{A,\pi^+\pi^-}=&e^2\frac{F_A^2}{M_A^2 f_\pi^2}\left\{s(\epsilon_1 \epsilon_2)-\frac{1}{4}
\frac{(\epsilon_1\epsilon_2)(t-m_\pi^2)^2-2 s(\epsilon_1 p_1)(\epsilon_2 p_1)}
{M_A^2-t-i0^+}
\right. \\
&\left. -\frac{1}{4}\frac{(\epsilon_1 \epsilon_2)
(u-m_\pi^2)^2-2s(\epsilon_1 p_1)(\epsilon_2 p_1)}{M_A^2-u-i0^+}\right\}~,
\end{aligned}
\label{expa}
\ea
and the same expression for $T_{A,K^+K^-}$ replacing $m_\pi\to m_K$. 
We use $M_A=1.23~\hbox{GeV}\simeq \sqrt{2}M_\rho$ because of Weinberg's sum rules \cite{weinsr}
 and the KSFR relation \cite{ksfr}. We do not coincide with the expression given in
ref.\cite{chino2} (also employed in ref.\cite{dh93}) for $T_{A,\pi^+\pi^-}$. However, we 
do agree with the formula given in ref.\cite{bgs94} for the similar 
exchange of the $1^{+-}$ axial-vector resonances, that has the same Lorentz structure.
In ref.\cite{chino2} the authors perform the check that at ${\cal O}(p^4)$ 
they reproduce  the $\chi$PT result for $\gamma\gamma\to\pi^+\pi^-$. 
This implies the axial-vector
saturation of the chiral counterterm $L_9+L_{10}$,
\be
L_9+L_{10}=\frac{F_A^2}{4M_A^2}~.
\label{satura}
\ee
 We also fulfill this condition since at ${\cal O}(p^4)$ our
expression is $T_{A,\pi^+\pi^-}=2e^2\frac{F_A^2}{M_A^2 f_\pi^2}(k_1 k_2)(\epsilon_1 \epsilon_2)+{\cal
O}(p^6)$ and agrees with that of ref.\cite{bc88} if eq.(\ref{satura}) is taking into account. 
 We use eq.(\ref{satura}) to fix $F_A$ from the value $L_9+L_{10}=(1.4\pm 0.3)\cdot 
10^{-3}$. In the following we shall replace $F_A^2$ by $L_9+L_{10}$ 
 using eq.(\ref{satura}). In ref.\cite{ximo} the couplings of the vector and axial-vector
 resonances to pseudoscalars and external sources are calculated within the 
 Extended
 Nambu-Jona-Laisinio model \cite{domenec}.

The S-wave projection of eq.(\ref{expa}) is,
\be
T_{A,P\bar{P}}^S=2e^2\frac{L_9+L_{10}}{f_\pi^2}\left[\frac{M_A^2}{\sigma_P}\log
\frac{1+\sigma_P+s_A/s}{1-\sigma_P+s_A/s}+s\right]~,
\ee
with $P=\pi^+$, $K^+$ and $s_A=2(M_A^2-m_P^2)$. This equation differs with the S-wave projection of
ref.\cite{dh93}, that employs the results of ref.\cite{chino2} which disagree with our
eq.(\ref{expa}). While we have the coefficient $M_A^2$ in front of the logarithm, 
ref.\cite{dh93} has $s_A/2$. This mistake was also performed in ref.\cite{oo98}, 
 because the expression from ref.\cite{dh93} was taken. 
 In addition, we have also included now the exchange of the $K^*$ vector resonance, 
 not taken into account in ref.\cite{oo98}. 
The numerical differences of correcting  for these
 latter deficiencies are however very small.

We finally give the expressions for the exchange of the $1^{+-}$ nonet 
of axial-vector resonances. We make use of the lowest order Lagrangian, see  
e.g. ref.\cite{bgs94}, 
\be
{\cal L}=\frac{e D}{f_\pi}F_{\mu\nu}\la \bar{C}^\mu \{Q,\partial^\nu \Phi\}\ra~,
\ee
where $C_\mu=\frac{1}{\sqrt{2}}C_\mu^i \lambda^i+\frac{1}{\sqrt{3}}C^9_\mu\cdot {\bf 1}$, with 
 standard notation. We then have,
\be
T_{C,P\bar{P}}=\frac{e^2 D^2}{18 f_\pi^2} \ell_P \left[\frac{(\epsilon_1 \epsilon_2)(t-m_P^2)^2
-2s(\epsilon_1 p_1)(\epsilon_2 p_1)}{t-M_C^2}+
\frac{(\epsilon_1 \epsilon_2)(u-m_P^2)^2-2s(\epsilon_1 p_1)(\epsilon_2 p_1)}{u-M^2_C}\right]~,
\ee
with $\ell_{\pi^0}=5$, $\ell_{\pi^+}=1/2$, $\ell_{K^+}=1/2$ and $\ell_{K^0}=2$. The S-wave
projection is 
\be
T_{C,P\bar{P}}^S=-\frac{e^2 D^2}{18 f_\pi^2}\ell_P
\left[s+\frac{M_C^2}{\sigma_P}\log\frac{1-\sigma_P+s_C/s}{1+\sigma_P+s_C/s}\right]~.
\ee
We take for $M_C=1.23$~GeV, the value corresponding to the mass of the $b_1(1235)$ resonance. 
The constant $D$ is determined from the $b_1\to \gamma \pi^+$ decay \cite{pdg},
\be
D^2=\frac{ M_C^3 \Gamma(b_1\to\gamma \pi^+)}{(M_C^2-m_\pi^2)^3}\frac{216 f_\pi^2}{\alpha}\simeq
3.2 \cdot 10^{-2}~.
\ee

\subsection{Isospin amplitudes}

Here we present our isospin amplitudes for $\gamma\gamma\to(\pi\pi)_I$. We employ the isospin states,
\ba
|\pi\pi\ra_0&=&\frac{-1}{\sqrt{3}}|\pi^+\pi^-+\pi^-\pi^++\pi^0\pi^0\ra~,\nn\\
|\pi\pi\ra_2&=&\frac{-1}{\sqrt{6}}|\pi^+\pi^-+\pi^-\pi^+-2\pi^0\pi^0\ra~,
\label{gordan}
\ea
with the subscript indicating the corresponding isospin. As a result, for the S-wave of
$\gamma\gamma\to(\pi\pi)_I$
\ba
T^S_{\gamma\gamma,0}&=&-\frac{2}{\sqrt{3}}\left(T^S_{\pi^+\pi^-}+\frac{1}{2}T^S_{\pi^0\pi^0}\right)~,
\nn\\
T^S_{\gamma\gamma,2}&=&-\sqrt{\frac{2}{3}} \left(
T^S_{\pi^+\pi^-}-T^S_{\pi^0\pi^0}\right)~.
\label{idecom}
\ea
This isospin decomposition applies not only to the whole S-wave amplitude  
but to all of its partial contributions, Born term, vector and axial exchanges, etc.

\section{Results}
\label{sec:resul}
In this section we discuss about the results that follow from the
use of the dispersive and dynamical approaches, 
sections \ref{sec:dis}
and \ref{sec:dyn}, respectively.

\subsection{Determination of the constants in the dispersive method}

Let us first consider the dispersive approach.
The $L_I(s)$ functions have  
contributions from the Born term and from the vector 
and axial-vector crossed exchanges:
\ba
L_0&=&-\frac{2}{\sqrt{3}}T_B^S-\frac{2}{\sqrt{3}}T_{A,\pi\pi}^S-
\sqrt{3}T_{\rho,\pi\pi}^S-\frac{1}{\sqrt{3}}T_{\omega,\pi\pi}^S+\frac{1}{3\sqrt{3}}T_{C,\pi\pi}^S~,\nn\\
L_2&=&-\sqrt{\frac{2}{3}}\left(T_B^S+T_{A,\pi\pi}^S-T_{\omega,\pi\pi}^S+\frac{1}{4}
T_{C,\pi\pi}^S\right)~,
\ea
where
\be
T_{C,\pi\pi}=\frac{e^2 D^2}{f_\pi^2}\left[s+\frac{M_C^2}{\sigma_\pi}
\log\frac{1-\sigma_\pi+s_C/s}{1+\sigma_\pi+s_A/s}\right]~.
\ee

We now turn to the coefficients $c_0$ and $c_2$ 
of eq.(\ref{f0f}). As discussed in section \ref{sec:dis}  these coefficients are 
determined by matching with the $\chi PT$ one loop results for the linear behaviour in $s$ of 
 $F_C(s)-B_C(s)$ and $F_N(s)$  when $s\to 0$. In order to impose the condition {\bf 1} one should realize that
 only $T_{A,\pi^+\pi^-}^S$ is ${\cal O}(s)$, being all the other resonance exchanges ${\cal O}(s^2)$. 
Then, 
\be
F_C(s)=T_B^S+4e^2 \chi \frac{L_9+L_{10}}{f_\pi^2} s-\frac{c_0 s}{\sqrt{3}}
-\frac{c_2 s}{\sqrt{6}}+{\cal O}(s^2)~,
\ee 
where $\chi=(1-m_\pi^2/2 M_A^2)/(1-m_\pi^2/M_A^2)$. The 
 one loop $\chi$PT result is 
\be
F_C(s)=T_B^S+4e^2\frac{L_9+L_{10}}{f_\pi^2}s+{\cal O}(s^2)~,
\ee
 and it follows that
\be
c_0+\frac{1}{\sqrt{2}}c_2= \xi 
\ee
with 
\be 
\xi = 4\sqrt{3}e^2\frac{L_9+L_{10}}{f_\pi^2}
\left(\chi-1\right)~.
\ee
 Notice that $\chi$ is very close to one and then 
$\xi\simeq 0$

 Regarding the condition {\bf 2},  the linear dependence in $s$ of $F_N(s)$ around $s=0$
  from eq.(\ref{f0f}) is given by  
\be
F_N(s)=-\frac{1}{\sqrt{3}}c_0 s+\sqrt{\frac{2}{3}}c_2 s+{\cal O}(s^2)~.
\ee
One loop $\chi$PT establishes that  
\be
F_N^S(s)=-e^2s/96\pi^2 f_\pi^2+{\cal O}(s^2)~,
\ee
 so that 
\ba
c_0&=&\frac{e^2 \sqrt{1/3}}{96\pi^2f_\pi^2}+ \frac23 \xi ~,\nn\\
c_2&=&-\frac{e^2\sqrt{2/3}}{96\pi^2 f_\pi^2} + \frac{\sqrt2}{3} \xi ~.
\label{cseqs}
\ea
We evaluate  the previous  expressions given by one loop $\chi$PT 
  either by employing $f_\pi=92.4$~MeV or $f\simeq 0.94 f_\pi$, where 
the former is the pion decay constant and the latter is the same but in the $SU(2)$ chiral limit 
 \cite{glchpt}. 
 This amounts around a $12\xpc$ of uncertainty in the evaluation of $c_0$ and $c_2$, due
 to the square dependence on $f_\pi$. 
Note that both choices, $f_\pi$ or $f$, are consistent with the 
 precision of the one loop calculation and the variation in
the results is an estimate for higher orders corrections.
 This is  included in the error analysis.

The constant $F_0(s_1)-L_0(s_1)$ in eq.(\ref{f0f}) is related with 
the  position of the Adler zero $s_A$, where $F_N(s_A)=0$, by  
\be
\begin{aligned}
\frac{\omega_0(s_A)}{\omega_0(s_1)}\frac{s_A^2}{s_1^2}\left[F_0(s_1)-L_0(s_1)\right]&=
\sqrt{2}L_2(s_A)-L_0(s_A)+\left( \sqrt{2}c_2\om_2(s_A)-c_0\om_0(s_A)\right)s_A \\
&+\sqrt{2}\frac{s_A^2}{\pi}\Omega_2(s_A)\int_{4m_\pi^2}^\infty 
\frac{L_2(s')\sin\phi_2(s')\,ds'}{{s'}^2(s'-s_A)|\om_2(s')|}\\
&-\frac{s_A^2}{\pi}\om_0(s_A)\int_{4m_\pi^2}^\infty\frac{L_0(s')\sin\bar\phi_0(s')\,ds'}
{{s'}^2(s'-s_A)|\om_0(s')|}~.
\end{aligned}
\ee
As discussed in the condition {\bf 3} of sec.\ref{sec:dis} 
the value of 
$F_0(s_1)-L_0(s_1)$ is finally fixed by requiring that 
$\sigma(\gamma\gamma\to\pi^0\pi^0)\leq 40$~nb at the $f_0(980)$ peak, at $\simeq s_1$. 
 This is based on the experimental data for $\gamma\gamma\to \pi^0\pi^0$ 
 of ref.\cite{crystal}, see  fig.\ref{fig:f0980} where the data including the 
$f_0(980)$ are shown,  whose  maximum value at the $f_0(980)$ peak of
$\sigma(\gamma\gamma\to\pi^0\pi^0)<40$ nb, errors included. In addition, we also take into account 
the resulting width  $\Gamma(f_0(980)\to
\gamma\gamma)=205^{+95}_{-83}(stat)^{+147}_{-117}(sys)$~eV  from the Belle Collaboration \cite{mori},
 and then impose that the width calculated from our approach lies within the interval
  $60$-$380$~eV, 
 as follows from the previous value. This interval is compatible with the one recently given for the 
 width $\Gamma(f_0(980)\to \gamma\gamma)$ in ref.\cite{penbelle} and with the value $\Gamma(f_0(980)\to
 \gamma\gamma)=0.2$~KeV predicted by the 
 dynamical approach \cite{oo98}.
 Note that  $\Gamma(f_0(980)\to \gamma\gamma)$ is proportional
to the maximum value of $\sigma(\gamma\gamma\to \pi^0\pi^0)$ at the $f_0(980)$ peak. For
practical purposes both constraints are equivalent.

\subsection{The $\gamma\gamma\to\pi^0\pi^0$ cross section}
 In our normalization, 
 the cross section in terms of $F_N(s)$ is given by
\be
\sigma(\gamma \gamma \longrightarrow \pi^0 \pi^0 ) = \frac{\sigma_\pi}{64 \pi s } |F_N(s)|^2~.
\ee

\begin{figure}[ht]
\centerline{\epsfig{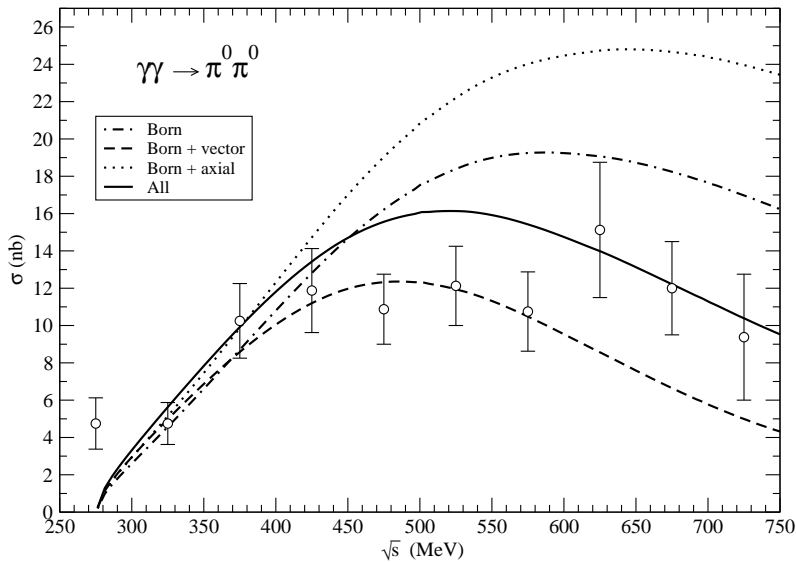}}
\vspace{0.2cm}
\caption[pilf]{\protect \small
 The $\gamma\gamma\to\pi^0\pi^0$ cross
section depending on the contributions included for $L_I(s)$. 
 Dot-dashed line (Born term only),   
 dashed line (Born term plus vector crossed exchanges) and 
dotted line (Born term plus axial-vector crossed exchange mechanisms).
 The solid line is the full result. 
\label{fig:BVA}}
\end{figure} 

 Since the main contribution in the low energy region to $\om_I(s)$ comes from
the low energy $\pi\pi$ phase shifts one needs to be rather precise  for the
 input parameterizations of the 
 $\pi\pi$ scattering data in this energy region in order to apply the dispersive method, 
eqs.(\ref{dispen}) and (\ref{f0f}). The small $I=2$ S-wave $\pi\pi$ phase
shifts, which induce small final state interaction corrections anyhow,
can be parameterized in simple terms and our fit compared to data can be seen in 
 fig.\ref{fig:phases}. For $I=0$  one of the parameterizations that we consider is
  the K-matrix of Hyams {\it et. al.} \cite{hyams},
\be
K_{ij}(s)=\alpha_i\alpha_j/(x_1-s)+\beta_i\beta_j/(x_2-s)+\gamma_{ij}~,
\label{km}
\ee
 where
 \be
 \begin{array}{lll}
 x_1^{1/2}=0.11\pm 0.15 & x_2^{1/2}=1.19\pm 0.01 & \\
\alpha_1=2.28\pm 0.08 & \alpha_2=2.02\pm 0.11 & \\
\beta_1=-1.00\pm 0.03 & \beta_2=0.47\pm 0.05 &\\
\gamma_{11}=2.86\pm 0.15 & \gamma_{12}=1.85\pm 0.18 & \gamma_{22}=1.00\pm 0.53~,
 \end{array}
 \label{eq:Hyamsparam}
 \ee
in units of appropriate powers of~GeV. 
The $K$ matrix is related to the $I=0$ S-wave  $T-$matrix by 
\begin{eqnarray}
T  &=& -8\pi\sqrt{s}\, (K^{-1} - i Q)^{-1}~,
\label{tmat}
\end{eqnarray}
where $Q=diag(q_\pi, q_K)$ with $q_\pi(q_K)$ the center of mass momentum of the pions (kaons). 
 The central values in eq.({\ref{eq:Hyamsparam}) correspond to 
  $\delta_\pi(s_K)_0<\pi$. 
  However, 
varying the parameters within their errors it is possible to obtain $\delta_\pi(s_K)_0>\pi$ 
 (see fig.\ref{fig:phases}   where both cases are shown).   
Since below 0.6~GeV this K-matrix was not fitted to data and strongly disagrees with 
$K_{e 4}$ data \cite{bnl,na48}, we consider instead 
the parameterizations of ref.\cite{cgl} (CGL), ref.\cite{py03} (PY) and 
  ref.\cite{alba} (AO).  
 They  agree with data from $K_{e4}$ decays 
\cite{bnl,na48} and span to a large extent the band of theoretical uncertainties in 
the $I=0$ S-wave $\pi\pi$ phase shifts
 \cite{hyams,kaminski,grayer}.  PY runs 
 through the higher values of $\delta_\pi(s)_0$, while CGL
  does through lower values, see fig.\ref{fig:phases}. AO, double-dash-dotted line, 
   moves somewhat in between the previous results
for energies between 0.5-0.7 GeV and then follows closely CGL results.
  CGL is used up to 0.8~GeV, since this is the upper limit of its analysis, and the
K-matrix eq.(\ref{km}) above that energy. On the other hand,  PY is employed up to 0.9~GeV, as in
this energy this parameterization agrees well inside errors with
\cite{hyams}, and  eq.(\ref{km}) above it. AO is used up to 1.5~GeV since it fits S-wave meson-meson
data up to 2 GeV.
 We shall also employ in some cases the results for the $\pi\pi$ phase shifts from 
ref. \cite{npa}, which gives rise to a $f_0(980)$ signal tested in many other processes like e.g.
$J/\Psi$ \cite{moo,jpsi}, $D$ \cite{ddecay}  and $\phi$ \cite{phi1,phi2} decays. 
 This parameterization will be referred in the following as OO and will be 
used in in the region $0.9\leq \sqrt{s}\leq 1.1$~GeV,  above it eq.(\ref{km}) is again taken
and below CGL.

\begin{figure}[ht]
\centerline{\epsfig{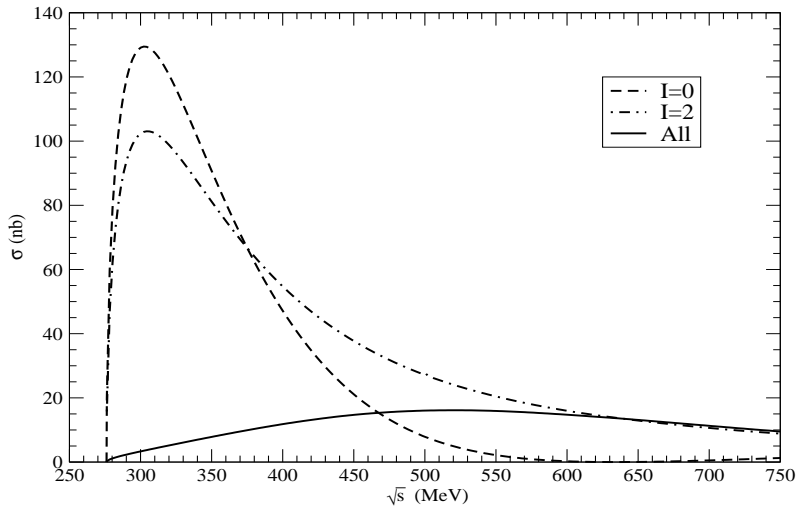}}
\vspace{0.2cm}
\caption[pilf]{\protect \small
$I=0$ (dashed) and 2 (dot-dashed) contributions to the 
$\gamma\gamma\to\pi^0\pi^0$ cross
section. The solid line is the full result. The curves have been generated employing the
parameterizations CGL and Hyams \cite{hyams}.
\label{fig:I02}}
\end{figure} 
 In fig.\ref{fig:BVA} we show the different contributions to the 
$\gamma\gamma\to\pi^0\pi^0$ cross section compared with 
data\footnote{Since the data only cover up to $|\cos\theta|<Z$,
with $Z=0.8$, we have divided them by the factor $Z$. This suited for low
energies as S-wave dominates, for higher energies is a normalization convention.} 
\cite{crystal}. The different curves originate by considering only the 
Born term as the only contribution to $L_I(s)$ in eq.(\ref{f0f}) (dot-dashed line),
 the Born term and the exchanges 
of vector resonances (dashed line), the Born term and the exchanges of axial-vector resonances
(dotted line) and the full model (solid line) for $L_I$, i.e., the Born term plus
vector and axial-vector resonance exchanges.  The curves have been
evaluated with the central values of  the parameterizations 
eq.(\ref{km}) and CGL. 
One can see that the dominant contribution comes from the dressing of the Born
term by final state interactions, despite that the Born term  at tree level is zero. 
 The vector and axial-vector resonance exchange
mechanisms also give important contributions
to the total cross section, even already at threshold for the exchange of the axial-vector resonances.
Most of the vector resonance crossed exchange contribution comes
from the $\omega$ exchange, since the $\omega$ coupling to
$\gamma\pi^0$  is about one order of magnitude larger than that due to 
 the $\rho$ or $K^*$ resonances.
It is clear from the figure that the axial-vector meson
exchange mechanisms,  overlooked in
 refs.\cite{penprl,penanegra}, are found to be substantial and tend
  to increase the cross section.
 Notice that the cross section given in ref.\cite{penprl}
 is very close  to the dashed line in fig.\ref{fig:BVA}. 
 
It is also worth stressing that there are strong
interferences between the different mechanisms, as
well as between the isospin $I=0$ and $I=2$ amplitudes.
In fig.\ref{fig:I02}, we show the $\gamma\gamma\to\pi^0\pi^0$
cross section considering only the $I=0$ S-wave
contribution  (dashed line),
the $I=2$ S-wave contribution (dot-dashed line) and the full result
(solid line).  For definiteness, the curves are calculated with the central value 
of ref.\cite{hyams} and CGL for low energies. 
We can see the strong cancellation between the 
$I=0$ and $I=2$ contributions due to the absence of the 
Born term at tree level in the  $\gamma\gamma\to\pi^0\pi^0$ channel, 
which is the dominant contribution in the $I=0$ and $I=2$ channels. 
 The Born term  is  the 
 responsible in every isospin channel for the bumps  
 at low energies in fig.\ref{fig:I02}, though the strong 
final state interactions in $I=0$ reshapes it considerably and tends to accumulate strength 
to lower energies than for the $I=2$ part.

\begin{figure}[ht]
\centerline{\epsfig{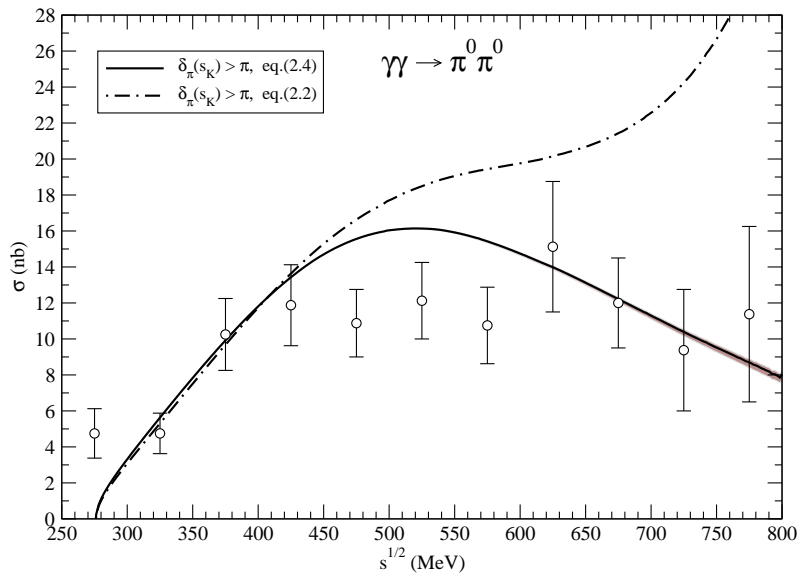}}
\vspace{0.2cm}
\caption[pilf]{\protect \small
The error band around the solid line 
 is the uncertainty in our results by requiring 
that $\sigma(\gamma\gamma\to \pi^0\pi^0)\lesssim 40$~nb at $s_1$. The solid line should be
compared with the dot-dashed one, that would correspond to the formalism of ref.\cite{penprl} in
the case that $\phi_0(s)\simeq \delta_\pi(s)_0$ above $s_K$ and including the
 axial-vector exchanges. The lines are evaluated with the central values of CGL and
 eq.(\ref{km}). \label{fig:updown} }
\end{figure}

In fig.\ref{fig:updown} we show with the solid line our results for the central values of eq.(\ref{km}) and 
CGL while the dashed-solid line corresponds to the result that would be obtained if using 
the approach of ref.\cite{penprl} (including also the axial-vector exchanges).
 The improvement achieved by 
passing from the dot-dashed line  to the solid one is determined by using eq.(\ref{f0f}) instead of 
 eq.(\ref{dispen}), being the latter the original equation derived
  in refs.\cite{penmorgan,penanegra,penprl}.  The gray band around the solid line
 corresponds to the uncertainty originated exclusively by taking
  $\sigma(\gamma\gamma\to\pi^0\pi^0)\leq 40$ nb at $s_1$.

\begin{figure}[ht]
\centerline{\epsfig{file=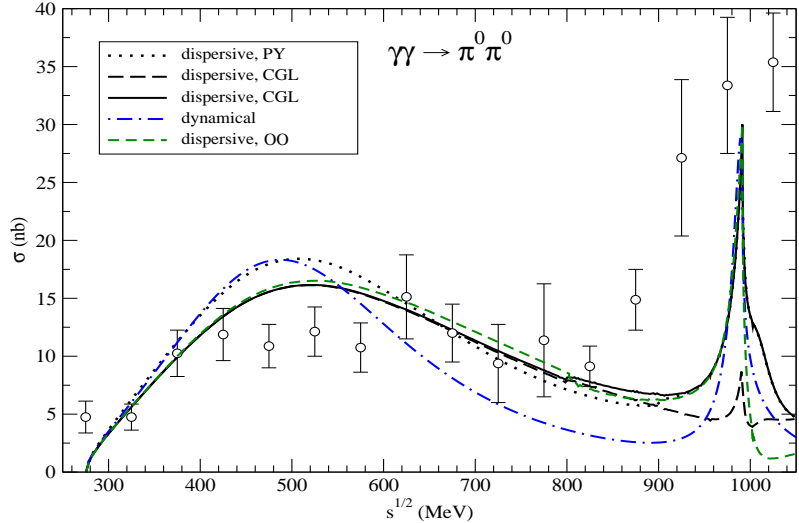,width=0.6\textwidth,angle=0}}
\vspace{0.2cm}
\caption[pilf]{\protect \small
 The $\sigma(\gamma\gamma\to \pi^0\pi^0)$ cross section including the region of the $f_0(980)$. 
 We show two different curves such that the cross
section at the top of the peak is 10 and 30~nb, long-dashed and solid lines, respectively. 
The short-dashed line employs the OO parameterization around the $f_0(980)$ resonance. 
The dot-dashed line is evaluated with the dynamical method of ref.\cite{oo98}. 
  \label{fig:f0980}}
\end{figure} 

We include in fig.\ref{fig:f0980} the $f_0(980)$ region as well, plotting for $\sqrt{s}\leq 1.05$~GeV.
 The generation of the $f_0(980)$ resonance peak within 
the dispersive method  is a remarkable achievement with important consequences as 
it has  been used already for fixing that the $z=+1$ behaviour of $\phi_0(s)$ is the one realized in 
nature. We have also used it to  constraint 
the values of $F_0(s_1)-L_0(s_1)$, so that we are able to 
take three subtractions in the dispersion relation instead of just two, as in previous studies. 
  The values of $F_0(s_1)-L_0(s_1)$ are fixed in this figure 
  such that the size of $\sigma(\gamma\gamma\to\pi^0\pi^0)$ in
 the $f_0(980)$ peak is  10 or 30~nb, dashed and solid lines, respectively. 
These lines are evaluated with the central values 
 of CGL and eq.(\ref{km}) for definiteness.  From this figure it is clear that  
 above $\sim 0.8$~GeV the  D-wave contribution in $\sigma(\gamma\gamma\to\pi^0\pi^0)$, due to the tail of
 the $f_2(1270)$, is important and this resonance is needed so as to agree width data. This contribution is added incoherently
because of the azimuthal integration (the $f_2(1270)$ resonance signal in $ \gamma\gamma$ has 
dominantly helicity 2 \cite{penmorgan}). We also show in the figure 
by the short-dashed line the result from the dispersive
approach but employing the parameterization OO for the $\pi\pi$ phase 
shifts around the $f_0(980)$ resonance.
 The dot-dashed line is calculated with the dynamical method of ref.\cite{oo98}. This method
rightly predicts the peak at $0.5$~GeV, with size very similar to that obtained from the dispersive 
method and using PY, though tends to give somewhat a too fast decreasing cross section 
for energies above $\gtrsim 0.6$ GeV. In fig.\ref{fig:fignew2} we also show the curve 
corresponding to the AO parameterization with a peak value in the $f_0(980)$ of 20 nb. In this way one
 can see how the actual shape of the $f_0(980)$ signal changes depending on the parameterizations 
 and on the peak values used. Further theoretical 
precision in this energy region requires to include the D-wave
 $\gamma\gamma\to\pi \pi$ contribution and fit simultaneously the recent data on $\gamma\gamma\to
 \pi^+\pi^-$ \cite{mori}. This is beyond the scope of the present study.

\begin{figure}[ht]
\centerline{\epsfig{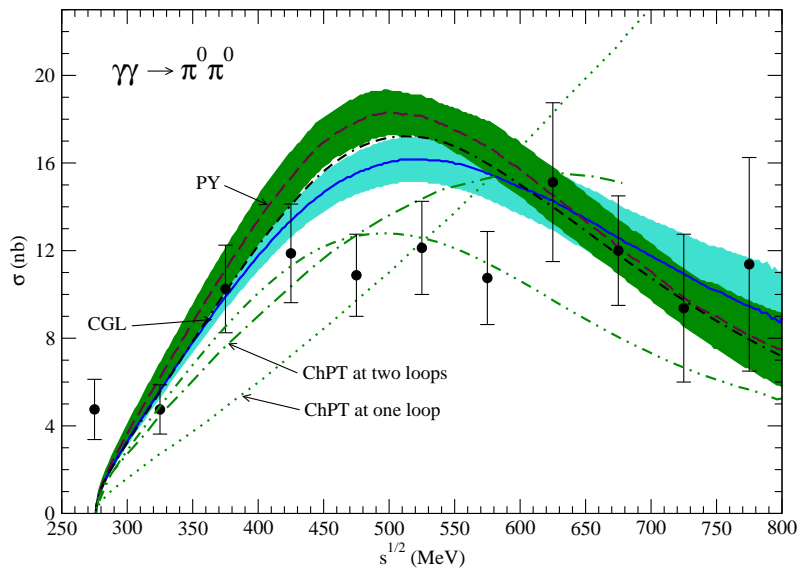}}
\vspace{-0.1cm}
\caption[pilf]{\protect \small
Final result for the $\gamma\gamma\to\pi^0\pi^0$ cross
section with the theoretical error band. The experimental data are from the Crystal Ball Collaboration
 ref.\cite{crystal}. The dashed line is evaluated with PY and the solid line with CGL. The band around each
 line corresponds to the theoretical uncertainties. The dot-dashed line corresponds to AO and does not
 include error band, of similar size than the ones shown, in order not to overload the figure. 
\label{fig:errorband}}
\end{figure}

Our final $\sigma(\gamma\gamma\to\pi^0 \pi^0)$ for $\sqrt{s}<0.8$~GeV 
 is shown in fig.\ref{fig:errorband}. We give
the results for CGL(solid) and PY(dashed) and eq.(\ref{km}), as usual. The corresponding band
around the lines has into account all the theoretical uncertainties
discussed with respect to the dispersive approach. For that, we implement a Monte-Carlo Gaussian sampling 
 taking into account the  errorbars in 
the parameters of CGL, PY and eq.(\ref{km}), the uncertainty for $s>s_K$ in the approximation 
$\phi_0(s)\simeq \delta_\pi(s)_0$ in the region $1\lesssim \sqrt{s}\lesssim 1.5$~GeV and 
that from eq.(\ref{asintotico}) above $s_H$.
 We also include in the error the $12\xpc$ uncertainty in evaluating 
$c_0$ and $c_2$ according to eq.(\ref{cseqs}) and the variation within the bound 
$\sigma(\gamma\gamma\to\pi^0\pi^0)\leq 40$ nb at $s_1$.  
 The width of the band is dominated by the errors in the parameterizations PY and CGL.
  In the same figure the dark dot-dashed line corresponds to the AO phase shifts. We do not show 
its corresponding error bar as it is of similar size as for CGL and PY and it would complicate the 
visualization of the information contained in the figure.
 In addition, the dotted line corresponds to one loop CHPT
 \cite{bc88,dhl87} and the lighter dot-dashed one is
 the two loop result \cite{bgs94,gis05}. The latter
  is closer to the dispersive results for low energies
   but still one observes that the ${\cal O}(p^8)$ corrections
 would be sizable. 
 It is worth stressing that if the axial-vector exchanges were removed,  in
 refs.\cite{penprl,penanegra} they were not included, then our curves would be smaller. This corresponds to the 
dot-dot-dashed line in fig.\ref{fig:errorband}, which is very close to that of 
 ref.\cite{penprl} when employing $\phi_0(s)\simeq \delta_\pi(s)_0-\pi$
for $s>s_K$. The former  curve is evaluated making use of CGL and ref.\cite{hyams} 
 with their central values. 
The three experimental points \cite{crystal}
 in the region $0.45$-$0.6$~GeV agree well with this curve.
However, once the axial-vector are included the curve rises. These three points lie around $1.5\sigma$ 
below the CGL result band, and by somewhat more than two sigmas below 
 the PY band. 
 This clearly shows that more precise experimental data on 
 $\gamma\gamma\to \pi^0\pi^0$  could  be used to distinguish between  
 different  low-energy   parameterizations of the $I=0$ $\pi\pi$ S-wave phase shifts. 
 The next three experimental points in the region $0.6$-$0.75$~GeV agree better when 
 the axial-vector resonance contributions are taken into account, as one should do. 
 As a result of this discussion, more precise experimental data for
 $\gamma\gamma\to\pi^0\pi^0$ are called for. The width of the
  error bands around the solid and dashed lines is reduced as compared with ref.\cite{orsletter} 
  because now the $z=+1$ case has been singled out.

\begin{figure}[ht]
\centerline{\epsfig{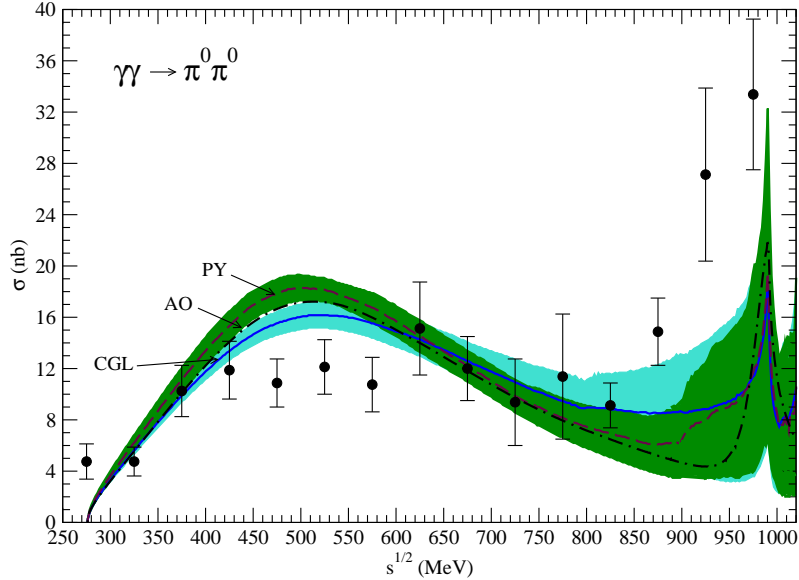}}
\vspace{-0.1cm}
\caption[pilf]{\protect \small
Final result for the $\gamma\gamma\to\pi^0\pi^0$ cross for $\sqrt{s}\leq 1.05$~GeV.
 The experimental data are from the Crystal Ball Collaboration
 ref.\cite{crystal}.   The dashed line is evaluated with PY and the solid line with CGL. The band around each
 line corresponds to the theoretical uncertainties. The dot-dashed line correspond to using AO. 
 Its error band, not shown, is similar to the ones of the other two curves.
\label{fig:fignew2}}
\end{figure}

In fig.\ref{fig:fignew2} we show our final results for  $\sigma(\gamma\gamma\to\pi^0\pi^0)$ including
the $f_0(980)$ region as well. The error bands are evaluated similarly as for fig.\ref{fig:errorband}. One
appreciates how the widths of the 
error bands largely increase in the $f_0(980)$ region because of the uncertainty in the
input for the value at the peak $4\leq \sigma(\gamma\gamma\to\pi^0\pi^0)\leq 40$~nb, which  
is obtained by requiring  that $\Gamma(f_0(980)\to\gamma\gamma)$, as calculated from our approach, 
lies in the range of the value determined in ref.\cite{mori}, 
$\Gamma(f_0(980)\to\gamma\gamma)=205^{+95}_{-83}(stat)^{+147}_{-117}(sys)$~eV. 
This bound in the size of the $f_0(980)$ peak could be also inferred by simple 
inspection of the experimental data in the same figure since otherwise the 
 $f_0(980)$ peak would be larger than the total cross-section, 
 which also includes incoherently a large
D-wave contribution from the tail of the $f_2(1270)$.
A reduction of this uncertainty
in the $f_0(980)$ has to await a proper study of the new data 
 $\gamma\gamma\to\pi^+\pi^-$ \cite{mori} along these lines, as already commented.

 From our approach we obtain that the Adler zero in $F_N(s)$ is located at   
 $s_A=(1.11\pm 0.03)~m_\pi^2$, if CGL is used, and at $s_A=(1.09\pm 0.04)~m_\pi^2$, when 
 PY is taken for lower energies.
  If AO is used up to 1.4~GeV we obtain for $s_A$ the same value  as the
 latter one. From our analysis then $s_A$ lies in the range $(1.05$-$1.14)~m_\pi^2$. This range 
 is in  agreement with the  the two loop $\chi$PT 
value \cite{bgs94}, $s_A=1.175~m_\pi^2$. The difference is around a $4\xpc$, well inside
 the expected uncertainties of ref.\cite{bgs94}.

\subsection{The $\sigma\to\gamma\gamma$ width}
\label{sec:sigwidth}

 Let us recall that the $\sigma$ is the lightest resonance with
the  quantum numbers of the vacuum. It was originally included
by studies of the $NN$ potential, but due to the advent of the chiral
potentials \cite{weinberg,epe,entem} it is no longer necessary and,
furthermore, chiral symmetry seems to substantially reduce its 
contributions \cite{osettoki}. Another relevant example where the $\sigma$ is introduced is 
 the linear sigma model \cite{aver}. However, when the non-linear sigma models
are unitarized, fulfilling the unitarity and analyticity requirements
associated with the right hand cut, a pole associated with the
$\sigma$ resonance appears as well \cite{iamcc,npa,nd,hannah,mixing,mrios}. 
The works \cite{npa,iamcc} were the first to show that the presence of such a light scalar resonance 
is not in contradiction with the non-linear sigma models. It appears because of the strong
interactions, driven by chiral symmetry,
 between the two pions in an $I=0$ S-wave pair. In ref.\cite{ollerkyoto} it was
shown that in the limit of chiral symmetry restoration, $f_\pi\to 0$, the
$\sigma$ pole moves to zero and becomes the chiral partner of the
pion. In this respect, the difference in mass between the  $\sigma$
resonance  and the $\pi$ is an order parameter for chiral symmetry
breaking. It is interesting to consider the $\sigma$ coupling to
$\gamma\gamma$ because, as indicated in the Introduction, it is
a complementary information apart from that of scattering as it is sensitive to the
electric charge. This can give clues to possible underlying
structures in its nature.

For the evaluation of the coupling  $\sigma\to \gamma\gamma$, $g_{\sigma\gamma\gamma}$, 
 the amplitude $F_N(s)$ has to be evaluated not on the first
or physical Riemann sheet but on the second one. This is
accomplished by shifting $q_\pi\to -q_\pi$.  We denote by $\widetilde{F}_N(s)$
the resulting amplitude on the second Riemann sheet. 
 The strong $I=0$ $\pi\pi$ S-wave $T-$matrix on the same sheet, $T_{II}(s)$, 
can be calculated in terms of the one on the physical  sheet, $T_I(s)$, by
\be
T_{II}^{-1}=T_I^{-1}-i\frac{q_\pi}{4\pi\sqrt{s}}~.
\label{t2nd}
\ee

 $T_{II}$ and $\widetilde{F}_N$ have a pole corresponding to the $\sigma$ resonance at 
$s_\sigma$, so that 
 \be
 \lim_{s\to s_\sigma}\widetilde{F}_N^S=
 \frac{g_{\sigma\gamma\gamma}\,g_{\sigma\pi^0\pi^0}}{s_\sigma-s}~,
\label{limit}
 \ee
where  $g_{\sigma\pi^0\pi^0}$ is 
the  coupling of the $\sigma$ to $\pi^0\pi^0$.\footnote{As it is well known \cite{weinbergbook} the phase of 
a resonance coupling is background dependent. Hence, the phase of $g_{\sigma\pi^0\pi^0}$ extracted 
from $\gamma\gamma\to\pi^0\pi^0$ differs from that obtained, e.g., from
$\pi^0\pi^0\to\pi^0\pi^0$. This does not bother us here since we are only interested in the 
moduli of $g_{\sigma\gamma\gamma}$ and $g_{\sigma\pi^0\pi^0}$.}

From $F_N$ of eq.(\ref{fcnpa}), corresponding to the extended dynamical approach 
of ref.\cite{oo98}, the coupling $g_{\sigma\gamma\gamma}$ is
\ba
\begin{aligned}
g_{\sigma\gamma\gamma}&=\sqrt{\frac{2}{3}}(\widetilde{t}_{A\pi^+\pi^-}
+1.1\widetilde{t}_{C\pi^0\pi^0}+\widetilde{t}_{\omega\pi^0\pi^0}
+2\widetilde{t}_{\rho\pi^+\pi^-}
+\widetilde{t}_{\chi \pi})g_{\sigma\pi\pi}\\
&+\frac{1}{\sqrt{2}}(\widetilde{t}_{AK^+K^-}+5\widetilde{t}_{CK^+K^-}
+2\widetilde{t}_{K^*K^+K^-}+\widetilde{t}_{\chi K})
g_{\sigma K\bar{K}}\\
&=g_{\sigma\pi\pi}\left[\sqrt{\frac{2}{3}}(\widetilde{t}_{A\pi^+\pi^-}+1.1\widetilde{t}_{C\pi^0\pi^0}
+\widetilde{t}_{\omega\pi^0\pi^0}
+2\widetilde{t}_{\rho\pi^+\pi^-}+\widetilde{t}_{\chi \pi})\right.\\
&\left.+\frac{1}{\sqrt{2}}(\widetilde{t}_{AK^+K^-}+5\widetilde{t}_{CK^+K^-}
+2\widetilde{t}_{K^*K^+K^-}+\widetilde{t}_{\chi K}) 
\frac{g_{\sigma K\bar{K}}}{g_{\sigma\pi\pi}}\right]
\end{aligned}
\label{gsigf}
\ea
In the previous equation $g_{\sigma\pi\pi}$ is the coupling
of the $\sigma$ to the $I=0$ $\pi\pi$
state in the unitary normalization of refs.\cite{oo98,mixing}\footnote{With the unitarity
normalization the width to $\pi\pi$ of an $I=0$ narrow resonance of mass $M$
 is given by $\Gamma=|g_{\sigma\pi\pi}|^2 q_\pi/(8\pi M^2)$.}
 and, analogously, $g_{\sigma K\bar{K}}$ is the coupling to the $I=0$ $K\bar{K}$ one. Notice 
that the ratio of couplings is given directly by 
$g_{\sigma K\bar{K}}/g_{\sigma\pi\pi}=t^0_{K\bar{K},K\bar{K}}/t^0_{\pi\pi,K\bar{K}}$, with the strong
amplitudes evaluated on the second Riemann sheet.

It is important to remark that the ratio  $g_{\sigma\gamma\gamma}/g_{\sigma\pi\pi}$  does not
 depend\footnote{Actually there is a dependence  from eq.(\ref{gsigf}) 
 on the strong coupling due to the kaons, but this term is very small numerically.}
 on the
strong meson-meson coupling, hence it allows a clearer comparison between
different  approaches for the $\gamma\gamma\to\pi\pi$ part.

We now determine $g_{\sigma\gamma\gamma}/g_{\sigma \pi\pi}$ 
and $\Gamma_{\sigma\to\gamma\gamma}$ by using the dispersive approach of
section \ref{sec:dis}.  The dispersion relation eq.(\ref{f0f}) to calculate $F_0(s)$ is only valid
on the first Riemann sheet. If evaluated on the second Riemann sheet there
would be an extra term due to the $\sigma$ pole. 
However, the relation between $F_0$ and
$\widetilde{F}_0$, the latter on the second sheet, can be easily
established using unitarity above the $\pi\pi$ threshold,
\be
F_0(s+i\epsilon)-F_0(s-i\epsilon)=-2i F_0(s+i\epsilon)\rho(s+i\epsilon)T_{II}^0(s-i\epsilon)~,
\label{f01}
\ee
 with  $4m_\pi^2<s<4m_K^2$, $\rho(s)=\sigma_\pi(s)/16\pi$ and $\epsilon\to0^+$.
  Due to the continuity when changing from one sheet to the other,
 \be
 F_0(s-i\epsilon)=\widetilde{F}_0(s+i\epsilon)~,~T_I^{I=0}(s-i\epsilon)={T}_{II}^{I=0}(s+i\epsilon)~.
\label{f02} \ee
 it results then from eq.(\ref{f01})
\be
\widetilde{F}_0(s)=F_0(s)\left(1+2i \rho(s)T_{II}^{I=0}(s)\right),
\label{analytic}
\ee
where the analytical extrapolation of eqs.(\ref{f01}) and 
eq.(\ref{f02}) has been taken. Around the $\sigma$ pole, 
\be
T_{II}^{I=0}=-\frac{g_{\sigma \pi\pi}^2}{s_\sigma-s}~,~\widetilde{F}_0(s)=\sqrt{2}\,\frac{g_{\sigma
\gamma\gamma}g_{\sigma\pi\pi}}{s_\sigma-s}~,
\label{coupres}
\ee
and then from eq.(\ref{analytic})
\be
g_{\sigma\gamma\gamma}^2=- g_{\sigma\pi\pi}^2\frac{1}{2}\left(\frac{\sigma_\pi(s_\sigma)}{8\pi}\right)^2 
F_0(s_\sigma)^2~.
\label{cdespejado}\ee
This expression allows to evaluate again the ratio 
$g_{\sigma\gamma\gamma}/g_{\sigma\pi\pi}$.

We consider in the subsequent $s_\sigma=(M_\sigma-i\Gamma_\sigma/2)^2$  either from the studies 
of ref.\cite{alba} or ref.\cite{caprini}. The former provides a reproduction of the S-wave $I=0$ and 
$K^-\pi^+\to K^-\pi^+$  data from $\pi\pi$ threshold up to around 2~GeV,
 considering for the first time 13 coupled channels for the $I=0$ meson-meson S-wave. 
The interactions kernels are derived from chiral Lagrangians and
 are unitarized employing the N/D method \cite{nd}.
Multiparticle states are mimicked by $\sigma\sigma$, $\rho\rho$, $\omega\omega$, $\ldots$. 
The $\sigma\sigma$
elementary transition amplitudes are given without any new free parameter,
 the same as for the vector-vector
ones which are derived by making use of a massive Yang-Mills theory \cite{ulfvecrev}. 
 In ref.\cite{alba} one has $M^{AO}_\sigma=(456\pm 6)$~MeV and $\Gamma^{AO}_\sigma=(482\pm 20)$~MeV. 
 Ref.\cite{caprini} provides the values  $M_\sigma^{CCL}=441^{+16}_{-8}~$MeV and 
$\Gamma_\sigma^{CCL}=544^{+18}_{-25}~$MeV. The latter values result from the $I=0$ $\pi\pi$ S-wave 
 amplitude obtained after solving the Roy
equations and matching with CHPT to two loops. 
In the following the superscripts $AO$ and $CCL$ refer to 
those results obtained by employing $s_\sigma$ from ref.\cite{alba} or \cite{caprini}, respectively.  
\begin{table}[ht]
\begin{center}
\begin{tabular}{|c|c|c|c||c|} 
\hline
$|g_{\sigma\gamma\gamma}/g_{\sigma\pi\pi}|\times 10^3$ & Born& Born + vector & Born + axial & All \\
\hline
Dispersive approach & $2.60$ & $2.22$ & $2.37$ & $2.01\pm 0.11$ \\ 
$s_\sigma$ from ref.\cite{caprini} & & & &\\
\hline
Dispersive approach & $2.42$ & $2.02$ & $2.25$ & $1.85\pm 0.09$ \\ 
$s_\sigma$ from ref.\cite{alba} & & & &\\
\hline
Dynamical approach & $2.05$ & $1.84$ & $1.81$ & $1.61$ \\ 
\hline
\end{tabular}
\end{center}
\caption{Different contributions to the ratio $|g_{\sigma\gamma\gamma}/g_{\sigma\pi\pi}|$ 
given by the dispersive approach of section \ref{sec:dis}, second and third row, and the 
dynamical one of section \ref{sec:dyn}, fourth row. In the second row the $s_\sigma$ pole position 
is taken from ref.\cite{caprini}, while for the third one $s_\sigma$  comes from ref.\cite{alba}.
 No error is given for the dynamical approach since this model is very
 constraint and the systematic uncertainties cannot be worked out.}
\label{tab:ratios}
\end{table}

In table~\ref{tab:ratios} we give $|g_{\sigma\gamma\gamma}/g_{\sigma\pi\pi}|$ for the dispersive and
dynamical approaches of sections \ref{sec:dis} and \ref{sec:dyn}, respectively. The second and 
third rows correspond to the dispersive approach with $s_\sigma$ from ref.\cite{caprini} and \cite{alba}, 
respectively. The result from the dynamical approach are shown in the last row. 
This approach makes 
use of strong amplitudes with $s_\sigma$ very close to that of ref.\cite{mixing}. 
 Notice that in order to calculate 
 $F_I(s)$, and then $\sigma(\gamma\gamma\to\pi^0\pi^0)$, $s_\sigma$ is not needed. 
 The different columns in table \ref{tab:ratios} represent different contributions to $L_I$. 
 From left to right in table \ref{tab:ratios}: 
only the Born term is taken as left hand cut, 
  Born term and vector resonance exchanges,
  Born term and axial-vector resonance exchanges and all together.
 We observe that the interferences between the different pieces has opposite 
 sign to those  in the $\gamma\gamma\to\pi^0\pi^0$ cross
section, fig.\ref{fig:BVA}. This difference occurs because in the
 $\pi^0\pi^0$ cross section both the $I=0$ and $I=2$ play a role, while
in the $\sigma$ width only the $I=0$ takes part. For the dispersive results we have taken the average 
the resulting ratios $|g_{\sigma\gamma\gamma}/g_{\sigma\pi\pi}|$ obtained with PY and CGL
parameterizations. The values obtained if using AO agree with the values shown 
within one sigma. No errorbar is given for the dynamical model result because this model is very
constraint, has no free parameters, and the systematic uncertainty cannot be estimated. Its
value is quite close to that of the second column, with $s_\sigma$ from \cite{alba},
only a 10$\%$ lower.

These ratios of residua at the $\sigma$ pole position 
 are the well defined predictions that follow from our $F_0(s)$, given $s_\sigma$. 
  However,  
  the radiative width to $\gamma\gamma$ for a wide resonance like the $\sigma$, 
  though more intuitive, has experimental 
  determinations that are parameterization dependent. This is  
  due to the non-trivial interplay between background and the broad resonant 
  signal as stressed in ref.\cite{orsletter}. An unambiguous
  definition is then required and we
  employ the standard narrow resonance width formula in terms of
  $g_{\sigma\gamma\gamma}$  as determined from the residue at $s_\sigma$,
\be
\Gamma(\sigma\to \gamma\gamma)=\frac{|g_{\sigma\gamma\gamma}|^2}{16\pi M_\sigma}~,
\label{sigmafoton}
\ee
as done in refs.\cite{penprl,orsletter}.
 Nevertheless, the determinations of the radiative widths  from this expression and those 
 from common experimental analyses of physical data
  can differ at the level of a $\lesssim 20\%$, as discussed 
 in ref.\cite{orsletter}. Notice that in the equations  usually employed in phenomenological
  fits to data, e.g. see
ref.\cite{mori}, the couplings are determined along the real axis, while the definition 
eq.(\ref{sigmafoton}) requires to move to the pole position.

The use of a narrow width resonance formula eq.(\ref{sigmafoton})
in order to evaluate $\Gamma(\sigma\to \gamma\gamma)$ is a precise definition and convenient 
criterion, though somewhat 
arbitrary as the $\sigma$ meson has a large width. 
 Elaborating a bit more on this issue let us introduce the normalized $\sigma$ mass
distribution $D(s)$,
\be
\int_{4m_\pi^2}^\infty D(s)ds=1~.
\label{nc}
\ee 
Then,
\be
\Gamma_{\sigma\to\gamma\gamma}=\frac{|g_{\sigma\gamma\gamma}|^2}{8\pi M_\sigma}\int_{4m_\pi^2}^\infty
\frac{q_\gamma}{s^{1/2}}D(s)~,
\label{ggg}
\ee
with $q_\gamma=\sqrt{s}/2$ is the photon center of mass three-momentum. 
 We can then rewrite eq.(\ref{ggg}) as,
\be
\Gamma_{\sigma\to\gamma\gamma}=\frac{|g_{\sigma\gamma\gamma}|^2}{16\pi M_\sigma}\int_{4m_\pi^2}^\infty
 D(s)ds=
\frac{|g_{\sigma\gamma\gamma}|^2}{16\pi M_\sigma}~.
\label{ggg2}
\ee
In the last step we have taken into account eq.(\ref{nc}). Despite the introduction of 
the mass distribution for the $\sigma$, $D(s)$, eq.(\ref{ggg2}) corresponds to
the narrow resonance width approximation due to the null mass of the photons.
 Of course, in the previous equations we have assumed 
 an energy independent coupling $g_{\sigma\gamma\gamma}$. A more physical calculation would imply to take
 this dependence into account but this is beyond the scope of the present paper. In the following, we shall
 content ourselves with eq.(\ref{sigmafoton}) as a well defined translation from 
 $|g_{\sigma\gamma\gamma}|$ to the more familiar concept of width.

In order to determine $\Gamma(\sigma\to\gamma\gamma)$ by applying eq.(\ref{sigmafoton}) with the values given in
table \ref{tab:ratios} one needs to provide numbers for $|g_{\sigma\pi\pi}|$. This is a crucial 
input to evaluate the width to two photons as the latter  is proportional to $|g_{\sigma\pi\pi}|^2$. 
 We first consider the value  $|g_{\sigma\pi\pi}^{AO}|=(3.17\pm 0.10)$~GeV from the 
 approach of ref.\cite{alba}.\footnote{We
  thank M. Albaladejo for supplying this number before publication.}
 The calculated width is then
\ba
\Gamma^{AO}(\sigma\to\pi\pi)&=&(1.50\pm 0.18)~\hbox{KeV}~.
\label{gammauchpt}
\ea

Not only the position of the pole in the partial wave amplitude, but also its residue 
can be calculated in the framework of the dispersive analysis described in ref.\cite{caprini}. 
Expressed in terms of the complex coefficient $g_{\sigma \pi \pi}$ defined 
in eq.(\ref{coupres}), the preliminary result for the residue amounts 
to $|g_{\sigma \pi \pi}| =(3.31^{+0.17}_{-0.08})$~GeV.\footnote{We 
express our gratitude to I. Caprini and H. Leutwyler for sending us this  
 result for the residue.}
\be
\Gamma^{CCL}(\sigma\to\gamma\gamma)=(1.98^{+0.30}_{-0.24})~\hbox{KeV}~,
\label{gammaccl}
\ee

Taking the average between the values in eqs.(\ref{gammauchpt}) and (\ref{gammaccl}) we end with,
\be
\Gamma(\sigma\to \gamma\gamma)=(1.68\pm 0.15)~\hbox{KeV}~.
\label{gammafinal}
\ee
This number is in agreement with the values $(1.8\pm 0.4)$~KeV, with $s_\sigma$ from
ref.\cite{mixing}, and $(2.1\pm 0.3)$~KeV, with $s_\sigma$ from ref.\cite{caprini}, calculated in 
ref.\cite{orsletter}. Note that  we use here the
most recent analysis of ref.\cite{alba} instead of ref.\cite{mixing} 
for the $\sigma$ pole position and coupling. In addition, the coupling 
$|g^{CCL}_{\sigma\pi\pi}|$ has been also explicitly evaluated from ref.\cite{caprini},
 which was not known at the time of refs.\cite{penprl,orsletter}. 
 On the other hand, eq.(\ref{gammafinal}) 
 agrees at the level of one sigma with the estimate 
 $\Gamma(\sigma\to \gamma\gamma)=(1.2\pm 0.4)~$KeV of ref.\cite{berni}, 
 that considers  the experimental data on the proton electromagnetic polarizabilities.

Pennington in ref.\cite{penprl} predicted $\Gamma^{CCL}(\sigma\to\gamma\gamma)=(4.09\pm 0.29)~$KeV 
for $s_\sigma$ from ref.\cite{caprini}. 
The difference between this number and our result in eq.(\ref{gammaccl}) is due to i)
ref.\cite{penprl}
calculates $|g_{\sigma\gamma\gamma}/g_{\sigma\pi\pi}|=(2.53\pm 0.09)\times 10^{-3}$ for $s_\sigma$ from
ref.\cite{caprini} instead of $(2.01\pm 0.11)\times 10^{-3}$ as given 
in the second row of table \ref{tab:ratios}. This is due
 to the omission of the axial-vector resonance
exchanges in ref.\cite{penprl,penanegra}, $10\%$, and to the improvement in our approach by
 taking into account 
one extra subtraction and slight different input for the phases, another $10\%$.
One then has the suppression factor $(2.01/2.53)^2=0.63$ for
the width $\Gamma(\sigma\to \gamma\gamma)$.  
ii) There is also the difference with regards the value 
of the $|g^{CCL}_{\sigma\pi\pi}|$ employed. Ref.\cite{penprl} uses
 $|g^{CCL}_{\sigma\pi\pi}|=3.86$~GeV
instead of $3.3$~GeV, with the last number very close to our 
 previous estimate of $|g^{CCL}_{\sigma\pi\pi}|$ in ref.\cite{orsletter}  (see eq.(3.14) of this 
 reference). 
The additional suppression factor $(3.3/3.86)^2=0.73$ then raises. iii) Finally, ref.\cite{penprl} 
includes and extra factor $|\beta(s_\sigma)|=0.956$ as compared with eq.(\ref{sigmafoton}). 
Taking all these several factors into account one has $4.09~\hbox{KeV}\to 4.09\times 0.63\times 0.73 \times
1.05=1.98$~KeV, as in eq.(\ref{gammaccl}). 

 As it is  stressed in ref.\cite{penprl},
 the $\sigma$ resonance is far from dominating the $\gamma\gamma\to\pi^0\pi^0$ reaction. Were this the
 case, then
 \be
 F_N(s)\simeq \frac{g_{\sigma\pi^0\pi^0}g_{\sigma\gamma\gamma}}{s_\sigma-s}~,
 \ee
 from where 
\be
\sigma(\gamma\gamma\to\pi^0\pi^0)\simeq \frac{8\pi}{3M_\sigma^2}
\frac{\Gamma(\sigma\to \gamma\gamma)}{\Gamma(\sigma\to\pi\pi)}~.
\ee
Given the value of $\sigma(\gamma\gamma\to\pi^0\pi^0)\simeq 16$~nb
in the peak, one then concludes that 
 the $\Gamma(\sigma\to\gamma\gamma)\simeq 0.39$~KeV, which is about
 a factor $4$ smaller than 
our calculation eq.(\ref{gammafinal}), and one order of magnitude smaller than 
the value of ref.\cite{penprl}.
 In this latter reference it was 
 argued that the reason for this  difference is the 
  very strong destructive interference between the
$I=2$ and $I=0$ amplitudes, see fig.\ref{fig:I02}. However,
  a quantitative analysis, like that offered in ref.\cite{penprl}, is affected by the 
fact that the ratio $|F_0/F_N|^2$ depends 
strongly on the precise 
energy where it is evaluated in the energy region where $\gamma\gamma\to\pi^0\pi^0$ 
peaks. This  is clearly seen in fig.\ref{fig:I02}. 
At 400 MeV the ratio $|F_0/F_N|^2\simeq 12$ (one has to
multiply by 3 the ratio of the cross sections 
of $I=0$ and $\pi^0\pi^0$ to correct for the Clebsch-Gordan
coefficient in eq.(\ref{gordan})), while  at 
500 MeV is  $\sim 3/2$.  This figure also shows that peak of the $\gamma\gamma\to\pi^0\pi^0$ 
cross section at around 500 MeV is not directly related to the $\sigma$ resonance, but a 
a result of the strong interference between the $I=2$ and $I=0$ components. Note also that 
the latter lacks of a
resonance structure around the $\sigma$ due to the Adler zero in $F_N(s)$.\footnote{See
refs.\cite{ddecay,rho} for a discussion on the role of the Adler zeroes 
and $\sigma$ resonant shapes in two pion production processes.}
 The $\sigma$ resonance, present in the $I=0$ component,
 distorts the form of the $I=0$ Born term   accumulating strength towards the threshold 
so that above 
 $\sqrt{s}\sim 0.5$~GeV $|F_0(s)|$ is much smaller than the $I=2$ counterpart.

\section{Conclusions}
\label{sec:conclu}
We have performed a theoretical study of the reaction $\gamma\gamma\to\pi^0\pi^0$ making use of 
 improved versions of the approaches of refs.\cite{orsletter} and \cite{oo98}. 
 The former is based on a dispersion
 relation for $\gamma\gamma\to \pi\pi$ in S-wave and definite isospin $I$ similar to that of 
 refs.\cite{penmorgan,penanegra,penprl}, but including  
 one more subtraction for $I=0$. This constant
  is fixed by taking into account an additional constraint 
based on a lax bound for the size of the $f_0(980)$ peak in $\gamma\gamma\to \pi^0\pi^0$ or, 
equivalently, for the value of $\Gamma(f_0(980)\to \gamma\gamma)$ from ref.\cite{mori}.
 In this way,
 ref.\cite{orsletter} was able to drastically reduce the uncertainty 
 in the results of ref.\cite{penprl} on   the function 
 phase $\phi_0(s')$ of $F_0(s)$ taken 
 above the $K\bar{K}$ threshold. 
Here, we have been able to  reduce it further so that it is shown that
 $\phi_0(s)$ should  follow quite closely $\delta_0(s)$ above the $K\bar{K}$ threshold. 
Then, in the advent of better experimental data on 
$\sigma(\gamma\gamma\to\pi^0\pi^0)$, our present results allow a sharper 
comparison with data  in order to
 distinguish between  different  parameterizations for the  low energy $\pi\pi$ $I=0$ S-wave. 
  
We have also updated the approach of ref.\cite{oo98} adding the  crossed 
exchanges of the $K^*$ and the nonet of the lightest 
 $1^{++}$ axial-vector resonances, not included in this reference. The axial-vector exchanges of the 
 $1^{+-}$ and $1^{++}$ resonances were not either taken into account in ref.\cite{penprl,penanegra}, but
 have been included here. 
 The explicit expressions and calculation details for the different
  elements both in the
 dispersive and dynamical approaches are also given. In particular, the expressions 
  for the Born terms and the crossed exchanges of the $1^{--}$,
 $1^{+-}$ and $1^{++}$ vector and axial-vector resonances, respectively,  are collected. 

The role of the $\sigma$ and $f_0(980)$ resonances in $\gamma\gamma\to\pi\pi$ has been also addressed. 
For the latter
resonance it was argued by continuity arguments
 that  it should come up as a small peak, as actually seen recently in ref.\cite{mori}
  for $\gamma\gamma\to\pi^+\pi^-$. Importantly, the
 fact that it is indeed a  peak requires that
  the phase of $F_0(s)$ should rapidly increased around the $K\bar{K}$ threshold, the so called 
  $z=+1$ scenario. 
Regarding the $\sigma$ resonance, we have calculated
 the ratio of the residua $|g_{\sigma\gamma\gamma}/g_{\sigma\pi\pi}|=1.91\pm 0.07$
   taking into account  the values of the $\sigma$ pole position, $s_\sigma$,
 of refs.\cite{caprini,alba}. 
 One should stress that  the calculation of $\sigma(\gamma\gamma\to\pi\pi)$ 
 is independent of the value taken for $s_\sigma$ and $|g_{\sigma\pi\pi}|$,
  though they are essential inputs  for calculating the width  
 $\Gamma(\sigma\to\gamma\gamma)$. With 
$|g^{AO}_{\sigma\pi\pi}|=(3.17\pm 0.10)$~GeV for $s_\sigma$ of 
ref.\cite{alba} one has $\Gamma^{AO}(\sigma\to\gamma\gamma)=(1.50\pm 0.18)$~KeV
 and with $|g^{CCL}_{\sigma\pi\pi}|=(3.31^{+0.17}_{-0.08})$~GeV, 
 a preliminary value corresponding to the $s_\sigma$ of ref.\cite{caprini},   
 $\Gamma^{CCL}(\sigma\to\gamma\gamma)=(1.98^{+0.30}_{-0.24})~\hbox{KeV}$. 
The average between the previous two values reads 
$\Gamma(\sigma\to\gamma\gamma)=(1.68\pm 0.15)~$KeV, that we take as our final result for the 
width. 
 
\section*{Acknowledgements}
 We thank Carlos Schat for fruitful discussions and 
 collaboration in this research  and in ref.\cite{orsletter} during his stay in the 
 Physics Department of the University of Murcia. 
  This work has been supported in part by the MEC (Spain) and FEDER (EC) Grants
  FPA2004-03470, FPA2007-62777 and Fis2006-03438,  the 
  Fundaci\'on  S\'eneca (Murcia) grant Ref. 02975/PI/05, the European Commission
(EC) RTN Network EURIDICE under Contract No. HPRN-CT2002-00311 and the HadronPhysics I3
Project (EC)  Contract No RII3-CT-2004-506078. 

\end{document}